\documentclass[11Pt]{article}
\usepackage{amssymb,euscript,latexsym,amsmath}
\usepackage{graphicx}
\usepackage[dvips]{epsfig}
\usepackage{color}
\usepackage{epstopdf}
\usepackage{setspace}
\usepackage{subfigure} 

\usepackage{caption2}

\DeclareGraphicsExtensions{.eps,.ps,.pdf}

\newcommand{\abs}[1]{\left| #1 \right|} 

\newcommand{\conj}[1]{\overline{#1}}
\newcommand{\real}[1]{\mathrm{Re} \left[ #1 \right]}
\newcommand{\imag}[1]{\mathrm{Im} \left[ #1 \right]}

\newcommand{\eee}[1]{\mathrm{e}^{ #1 }}
\newcommand{\ii}{\mathrm{i}} 

\begin{document}

\title{\Large \bf Dipole interactions in doubly-periodic domains}


\author{\normalsize  Alan Cheng Hou Tsang and Eva Kanso \\[2ex]
{\footnotesize Aerospace and Mechanical Engineering, University of Southern California} \\
{\footnotesize 854 Downey Way, Los Angeles, CA 90089-1191} \\
}

\maketitle

{\small

\abstract{We consider the interactions of  \textit{finite dipoles} in a doubly-periodic domain. 
A finite dipole is a pair of equal and opposite strength point vortices separated by a finite distance. 
The dynamics of multiple finite dipoles in an unbounded inviscid fluid was first proposed by
Tchieu, Kanso \& Newton in~\cite{Tchieu:prsa2012a} 
as a model that captures the ``far-field" hydrodynamic interactions in fish schools. 
In this paper, we formulate the equations of motion governing the dynamics of finite-dipoles in 
a doubly-periodic domain. We show that a single dipole in a doubly-periodic domain exhibits periodic 
and aperiodic behavior, in contrast to a single dipole in an unbounded domain.
 In the case of two dipoles in doubly-periodic domain, 
we identify a number of interesting trajectories including collision, collision avoidance, 
and passive synchronization of the dipoles. We then examine two types of dipole lattices: 
rectangular and diamond. We verify that these lattices are in a state of relative equilibrium and
 show that the rectangular lattice is unstable while the diamond lattice is linearly stable for a range 
of perturbations. We conclude by commenting on  the insights these models provide in 
the context of fish schooling.}
}

\doublespacing

\section{Introduction}

\begin{figure}[!t]
\begin{center}
\includegraphics[width=\textwidth]{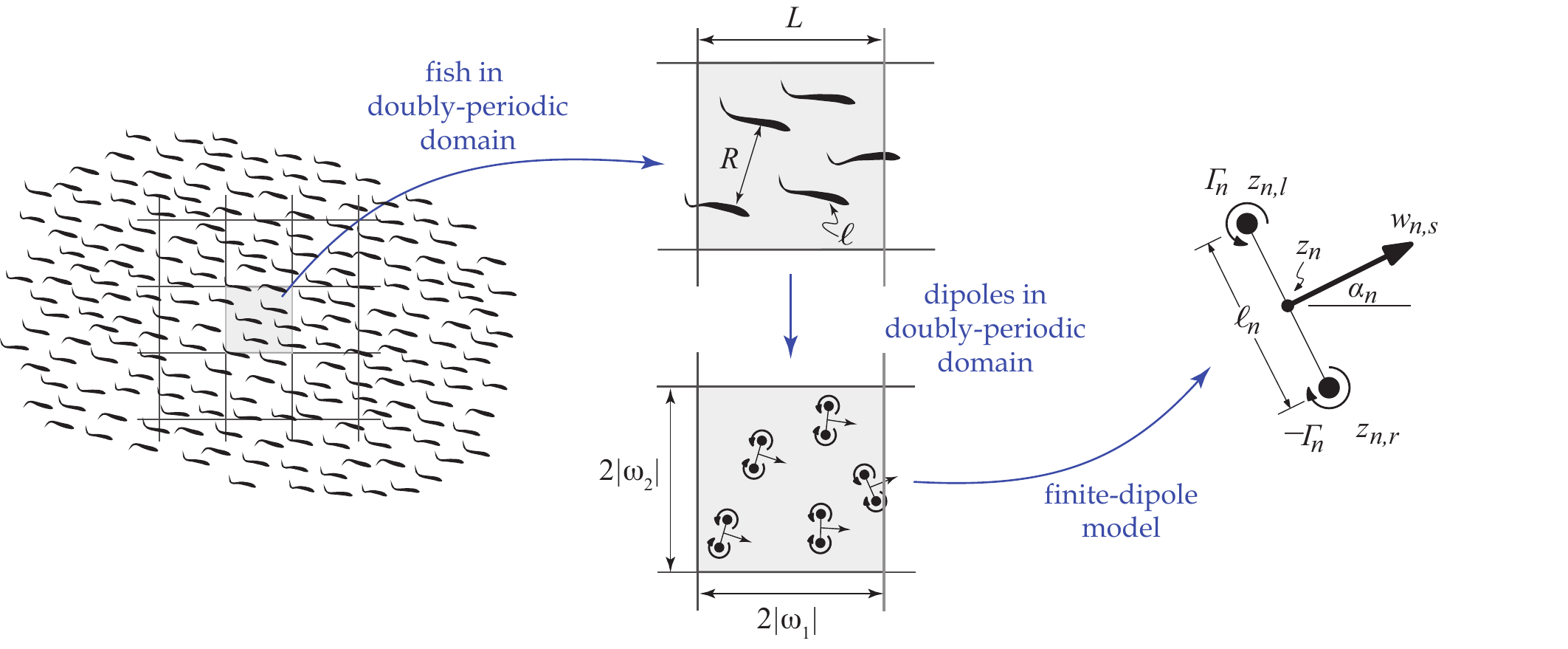}
\caption{\footnotesize Schematic showing a homogeneous school of fish $($left$)$. To highlight fish interactions in a domain within the school as opposed to on its boundary, we use fish in a domain with periodic boundary conditions $($middle$)$. 
Individual fish are modeled using the finite dipole model $($right$)$.}
  \label{fig:schematic}
  \end{center}
\end{figure}

The question of how interactions among individual fish result in highly-coordinated motion in fish schools has been the focus of numerous studies, the majority of which consider behavior-based models of homogeneous particles (fish) 
interacting locally based on rules of repulsion, alignment and attraction to other fish (see, for example,~\cite{couzin:jtb2002a} and~\cite{Couzin:n2005a}). These models are 
capable of exhibiting realistic dynamics similar to those observed in biological schools (see~\cite{Parrish:bb2002a}), but do not elucidate the mechanisms by which individuals transmit and integrate information from the school to guide their motion as noted in~\cite{Katz:pnas2011a}. 
In particular, little is known about the role of the fluid medium in guiding the motion of the individual 
fish. Our main motivation in this paper is to develop a framework for studying the hydrodynamic interactions inside a 
large school of fish, as opposed to near the school boundary, see Figure~\ref{fig:schematic}(left). We assume the school is homogeneous and we focus on fish interactions in a domain within the school. More specifically, we consider fish interactions in a rectangular domain with periodic boundary conditions. This argument holds when 
the characteristic length $L$ of the domain is small relative to the school size but much larger than the fish size $\ell$ and the separation distance $R$ between two neighboring fish.

As a leading-order model of the fish motion, we use $N$ finite dipoles in a doubly-periodic 
domain, see Figure~\ref{fig:schematic}(middle). A finite dipole is a pair of equal and opposite 
strength point vortices separated by a finite distance, see Figure~\ref{fig:schematic}(right), 
and thus its self-propelled speed is well-defined as opposed to that of a point dipole. 
For point dipole models, see~\cite{newton:dcds2005a}, \cite{yanovsky:pl2009a}, \cite{kulik:tmp2010a} and \cite{smith:pnp2011a}. 
Another reason for using the finite-dipole model is that a body propelling itself in a two-dimensional 
inviscid fluid produces, to leading order, a dipolar velocity field. For several hydro-dynamically 
coupled swimming bodies, when their separation distance $R$ is large relative to their size $\ell$, 
\cite{Tchieu:prsa2012a} argued that a dynamical system consisting of a collection of interacting 
{finite-sized} self-propelled dipoles would be a reasonable model governing the far-field hydrodynamic 
interactions among the bodies. Numerical evidence in~\cite{Tchieu:prsa2012a} suggests that the finite 
dipole model is a good approximation of interacting bodies even when the separation distance $R\sim 3\ell$ 
is of order $\ell$.

In this paper, we generalize the formulation of~\cite{Tchieu:prsa2012a} to the case of a doubly-periodic domain.
Given that each finite dipole consists of two constrained point vortices, we build upon known results
on vortex interactions in doubly-periodic domains. The basic formulation of a simple vortex lattice, which is a special 
 case of a vortex in a doubly-periodic domain, was first discussed by~\cite{tkachenko:sjetp1966a} and was extended by~\cite{oneil:jmp1989a} to general lattices. Vortices in periodic and doubly-periodic domains were further studied in~\cite{aref:jfm1996a},~\cite{Stremler:jfm1999a} and~\cite{Stremler:tcfd2010a}. Clusters of point vortices in doubly-periodic domains were also considered in~\cite{umeki:jpsj2007a}. 

The organization of this paper is as follows. Section \ref{sec:formulation} addresses the formulation of $N$ finite dipoles in a doubly-periodic domain as a constrained $2N$ point vortex system. This is done by directly modifying the
 standard point vortex equations of motion to respect the constraint that each pair of point vortices of equal and opposite  strength remain a fixed distance $\ell$ apart. Section~\ref{sec:singledipole} discusses the periodic and aperiodic behavior  of one dipole in a doubly-periodic domain. Examples of interactions of two finite dipole systems are presented in section \ref{sec:interactions}. Section \ref{sec:formation} focuses on two types of dipole lattices: rectangular and diamond. We show that both lattices are in a state of relative equilibrium with the former being unstable while the latter is linearly stable to a range of small perturbations. We conclude by commenting on the insights of these models offer in the context of schooling of fish.


\section{Problem Formulation}
\label{sec:formulation}

Consider $N$ pairs of point vortices or {\em dipoles} of equal and opposite strengths ($\pm\Gamma_n$) 
placed a distance $\ell_{n}$ apart, $n=1,\ldots, N$. See Figure~\ref{fig:schematic}(middle) for a depiction
of $N$ dipoles in a doubly-periodic domain and Figure~\ref{fig:schematic}(right) as well as
for the details of a single finite dipole.  The vortex of strength $+\Gamma_{n}$ is referred to as 
the left vortex and its position is denoted by $z_{n,\textrm{l}}$ whereas the $-\Gamma_n$ vortex is called 
the right vortex and its position is denoted by  $z_{n,\textrm{r}}$. 
For convenience, complex notation ($z=x+\ii y$ and $\ii=\sqrt{-1}$) is employed. 
The position $z_n$ of the dipole center is related to $z_{n,\textrm{l}}$ and $z_{n,\textrm{r}}$ via
\begin{equation}
	\label{eq:formulation:dipoleLocation}
	z_{n} = \frac{z_{n, \mathrm l}+z_{n,\mathrm r}}{2}.
\end{equation}
Let $\alpha_{n}$ represents the orientation of the dipole with respect to the $x$--axis. Then, the position
of the left and right vortices is given by,
\begin{equation} \label{eq:formulation:vortexLocation}
\begin{split}
    z_{n,\textrm{l}} = z_{n} + \frac{\ii \ell_{n} \eee{\ii \alpha_{n}}}{2}, \qquad
		z_{n,\textrm{r}} = z_{n} - \frac{\ii \ell_{n} \eee{\ii \alpha_{n}}}{2},
\end{split}
\end{equation}

Our goal is to formulate the equations of motion governing the interaction between $N$ dipoles in a doubly-periodic domain while the vortex pair in each dipole is constrained to have a finite length $\ell_n$  ($\dot{\ell}_{n} = 0$), hence the name \textit{finite dipole}. This amounts to deriving equations of motion for all dipole centers $z_{n}$ and the orientations $\alpha_{n}$. Following~\cite{Tchieu:prsa2012a}, we assume that the constraint $\dot{\ell}_{n} = 0$ induces an additional inter-dipole velocity to the $2 N$-vortex problem. That is to say, we assume that the left and right vortices are advected according to the velocity
\begin{subequations} 
	\label{eq:formulation:velocityVortex}
  \begin{align}
    \dot{\conj{z}}_{n, \textrm l} = w_{n,\mathrm{s}} + w_{n,\mathrm o} (z_{n,\mathrm l}) + \ii \lambda_{n} \eee{- \ii \alpha_{n}}, \\
		\dot{\conj{z}}_{n, \textrm r} = w_{n,\mathrm{s}} + w_{n,\mathrm o} (z_{n,\mathrm r}) - \ii \lambda_{n} \eee{- \ii \alpha_{n}},
  \end{align}
\end{subequations}
where $\conj{( \ )}$ and $\dot{ (\ )}$ represent the complex conjugate and time derivative, respectively, and $\lambda_{n}$ is a real constant. The term $w_{n,\mathrm{s}}$ represents the sum of self-induced conjugate velocity of each dipole $n$ and conjugate velocity induced by the dipole's own images. Note that due to the doubly-periodic nature of the domain, each dipole has infinitely many images. The term $w_{n,\mathrm{o}}$ represents the conjugate velocity induced by all other finite dipoles and their images. 

The term $\pm \ii \lambda_{n} \eee{- \ii \alpha_{n}}$ in~\eqref{eq:formulation:velocityVortex} is an additional, attractive ($\lambda_{n} > 0$) or repulsive ($\lambda_{n} < 0$) inter-dipole velocity that allows us to apply the finite-length constraint on $\ell_{n}$.  Physically speaking, this term forces the two vortices $z_{n,\mathrm l}$ and $z_{n, \mathrm r}$ to stay a fixed distance apart by allowing them to overcome the tendency to move toward or away  from each other with speed $|\lambda_{n}|$ along the line joining these two vortices. The introduction of non-zero $\lambda_{n}$ can be thought of as introducing a degree of freedom to enforce the constraint that $\dot{\ell}_{n} = 0$, much like applying Lagrange multipliers in constrained mechanics. This allows each finite dipole to retain its `particle-like' identity throughout its time evolution. It is noted in~\cite{Tchieu:prsa2012a} that $\lambda_n$ 
translates into constraint forces acting on the Euler equation governing the fluid system, thus breaking the Hamiltonian nature of the system.

Equations~\eqref{eq:formulation:velocityVortex} can be rewritten as a system of equations governing the motion of the center $z_n$ and the orientation $\alpha_n$ of each finite dipole. This is done exactly as in \cite{Tchieu:prsa2012a} for the unbounded plane except that $\omega_{n,s}$ and $\omega_{n,o}$ have different expressions in the doubly-periodic domain as will be shown below. Namely, upon substituting~\eqref{eq:formulation:vortexLocation} into~\eqref{eq:formulation:velocityVortex}, one gets
\begin{align}
	\label{eq:formulation:eomZn}
	\dot{\conj{z}}_{n} = w_{n,\mathrm{s}} + \frac{w_{n,\mathrm o}(z_{n,\textrm l}) + w_{n, \mathrm o}(z_{n,\textrm r})}{2},
\end{align}
\begin{equation}
	\label{eq:formulation:eomAlphaN}
	\dot{\alpha}_{n} =  \frac{ \real{\left( w_{n,\mathrm o} (z_{n, \mathrm r }) - w_{n, \mathrm o} (z_{n, \mathrm l}) \right) {\eee{\ii \alpha_{n}}}}}{\ell_{n}},
\end{equation}
and
\begin{equation}
	\label{eq:formulation:eomLambdaN}
	\lambda_{n} = \frac{1}{2} \imag{ \left( w_{n,\mathrm o} (z_{n,\mathrm l}) - w_{n, \mathrm o} (z_{n,\mathrm r}) \right) \eee{\ii \alpha_{n}}}.
\end{equation}
In order to close this system of equations, one needs to find expressions for the self-induced velocity $w_{n,\mathrm{s}}$  
and the velocity  $w_{n,\mathrm{o}}$ induced by other dipoles in a doubly-periodic domain.

To obtain expressions for the terms $w_{n,\mathrm{s}}$ and $w_{n,\mathrm{o}}$,  first remember that the velocity field created by {\em unconstrained} vortices in a doubly-periodic domain is given in terms of the Weierstrass-$\zeta$ function (see, for example,~\cite{tkachenko:sjetp1966a}, \cite{Stremler:jfm1999a}, \cite{Stremler:tcfd2010a} and \cite{umeki:jpsj2007a}).
This yields, upon straightforward manipulations, that the conjugate velocity field created by the 
\textit{unconstrained} but paired $2N$ point vortices in a doubly-periodic domain is given by
\begin{equation}
\begin{split}
\label{eq:formulation:velocityDipolePeriodic}
	  \dot{\conj{z}} & =\frac{\Gamma_{n}}{2\pi\ii} \left[\zeta\left(z-z_{n,l};\omega_{1},\omega_{2}\right)-\zeta\left(z-z_{n,r};\omega_{1},\omega_{2}\right) \right.
	   \\[2ex] 
	  & \hspace{1.75in}  +\left(\frac{\pi\conj{\omega}_{1}}{\Delta\omega_{1}}-\frac{\eta_{1}}{\omega_{1}}\right)\left(z_{n,r}-z_{n,l}\right)
	  -\frac{\pi}{\Delta}\left(\conj{z_{n,r}}-\conj{z_{n,l}}\right)\left.\right].
\end{split}
\end{equation}
where $\zeta(z;\omega_{1},\omega_{2})$ is the Weierstrass $\zeta$-function, 
\begin{equation}
	\label{eq:formulation:WeierstrassZeta}
	\begin{split}
	\zeta\left(z;\omega_{1},\omega_{2}\right)=\frac{1}{z}+\sum_{p,q} \frac{1}{z-\Omega_{pq}} +\frac{1}{\Omega_{pq}}+\frac{z}{\Omega_{pq}^{2}}. \\
	\Omega_{pq}=2p\omega_{1}+2q\omega_{2},	\qquad
	p,q\in \mathbb{Z} \! - \!\{0\}. 
	\end{split}
\end{equation}
Here, $p$ and $q$ are signed integers. In~\eqref{eq:formulation:velocityDipolePeriodic},  $\omega_{1}$ and  $\omega_{2}$ are the half-periods of the doubly-periodic domain, 
$\eta_{1}$ is the value of the Weierstrass $\zeta$-function at the half-period $\omega_{1}$, and $\Delta$  is the area of the rectangular domain and is given by the Legendre's relation, 
$\Delta=2\ii(\omega_{1}\conj{\omega}_{2}-\conj{\omega}_{1}\omega_{2})$. Note that the third term in~\eqref{eq:formulation:velocityDipolePeriodic} goes to zero for a square domain
 ($\omega_{1}$ is purely real and $\omega_2$ is purely imaginary, such that $\omega_{1}$ = $\abs{\omega_{2}}$) ). 

By virtue of~\eqref{eq:formulation:velocityDipolePeriodic}, 
one can readily verify that $w_{n,\mathrm{s}}$ takes the form
\begin{equation}
	\label{eq:formulation:wnsperiodic}
	w_{n,\mathrm{s}} = \frac{\Gamma_{n}}{2\pi\ii}\left[\zeta\left(-\ii \ell_{n} \eee{\ii \alpha_{n}}\right)-\left(\frac{\pi\conj{\omega}_{1}}{\Delta\omega_{1}}-\frac{\eta_{1}}{\omega_{1}}\right)\ii \ell_{n}\eee{\ii \alpha_{n}}-\frac{\pi}{\Delta}\ii \ell_{n}\eee{-\ii \alpha_{n}}\right] ,
\end{equation}
whereas the term $w_{n,\mathrm o}$ representing the velocity induced by all other dipoles and their images is given by
\begin{equation}
\begin{split}
\label{eq:formulation:wnoperiodic}
	  w_{n,\mathrm o} (z) & =  \sum_{j\neq n}^N\frac{\Gamma_{j}}{2\pi\ii} \left[\dfrac{}{}\zeta\left(z-z_{j,l};\omega_{1},\omega_{2}\right)
	  \right.  -\zeta\left(z-z_{j,r};\omega_{1},\omega_{2}\right)
	 \ - \\[0.5ex]
	 &\hspace{1.75in} \left. \left(\frac{\pi\conj{\omega}_{1}}{\Delta\omega_{1}}-\frac{\eta_{1}}{\omega_{1}}\right)\ii \ell_{j} \eee{ \ii \alpha_{j}}
	  -\frac{\pi}{\Delta}\ii \ell_{j} \eee{ -\ii \alpha_{j}}\right].
\end{split}
\end{equation}
Equations~\eqref{eq:formulation:wnsperiodic}, \eqref{eq:formulation:wnoperiodic} can now be substituted back 
into~\eqref{eq:formulation:eomZn}, \eqref{eq:formulation:eomAlphaN}
to get a closed system of $3N$ real equations ($N$ complex + $N$ real) governing the motion of $N$ finite dipoles interacting in a doubly-periodic domain. 
The strength of the Lagrange multiplier $\lambda_n$ is obtained by substituting \eqref{eq:formulation:wnsperiodic}, \eqref{eq:formulation:wnoperiodic}  into~\eqref{eq:formulation:eomLambdaN}.

Two remarks are in order here. First, when the period of the system goes to infinity, $\omega_1,\omega_2 \rightarrow\infty$, the self-induced velocity and the velocity 
induced by other dipoles are reduced to their counterparts in the case of an unbounded plane, namely,
\begin{equation}	\label{eq:formulation:wnsi}
\begin{split}
	w_{n,\mathrm{s}} & = \frac{\Gamma_{n} \eee{-\ii \alpha_{n}}}{2 \pi \ell_{n}}, \\[2ex]
	w_{n,\mathrm o} (z) & =\sum_{j\neq n}^N \frac{\Gamma_j}{2 \pi \ii} \left( \frac{1}{z -z_{j,\textrm l}} - \frac{1}{z-z_{j,\textrm r}} \right).
	\end{split}
\end{equation}
Second, in the {\em unconstrained} interaction of $N$ dipoles (or equivalently, $2N$ point vortices) in a doubly-periodic domain, the inter-dipole spacing $\ell_{n}$ is not constant 
and the equations of motion for each vortex are given by substituting \eqref{eq:formulation:wnsperiodic}, \eqref{eq:formulation:wnoperiodic} into \eqref{eq:formulation:velocityVortex} with $\lambda_{n} = 0$.  
The system of equations is then Hamiltonian and the total linear impulse is conserved. 
The Hamiltonian of a $2N$-vortex system subject 
to periodic boundary conditions can be written as
\begin{equation}
\begin{split}
\label{eq:formulation:hamVortexPeriodic}
	  	H& =-\frac{1}{4\pi} \sum_{n=1}^{2N}\sum_{j=1}^{2N}\Gamma_{n}\Gamma_{j}  \Bigl\{ \Bigr.  \ln\abs{\sigma(z_n-z_j)} \\[1ex]
		& \hspace{1.25in} +\mathrm{Re}\left[\left(\frac{\pi\conj{\omega}_{1}}{\Delta\omega_{1}}-\frac{\eta_{1}}{\omega_{1}}\right)\frac{(z_n-z_j)^2}{2}\right]-\frac{\pi}{2\Delta}[(x_n-x_j)^2+(y_n-y_j)^2] \Bigl. \Bigr\}.
\end{split}
\end{equation}
where ${n}\neq{j}$, $\sigma(z)$ is the Weierstrass sigma function and $\zeta(z)$ is the logarithmic derivative of $\sigma(z)$. The total linear impulse of a $2N$-vortex system $\sum_{j=1}^{2N}\Gamma_{j}z_{j}$ is conserved, whereas the angular impulse $\sum_{j=1}^{2N}\Gamma_{j}z_{j}\conj{z}_{j}$ is not conserved in periodic domains.
This Hamiltonian structure is destroyed in the finite dipole system due to the constraint $\dot{\ell}_n=0$ and the associated Lagrange multiplier $\lambda_n$ as noted above.


\section{Periodic and aperiodic behavior of single dipole}
\label{sec:singledipole}

\begin{figure}[!t]
\begin{center}
\includegraphics[scale=0.5] {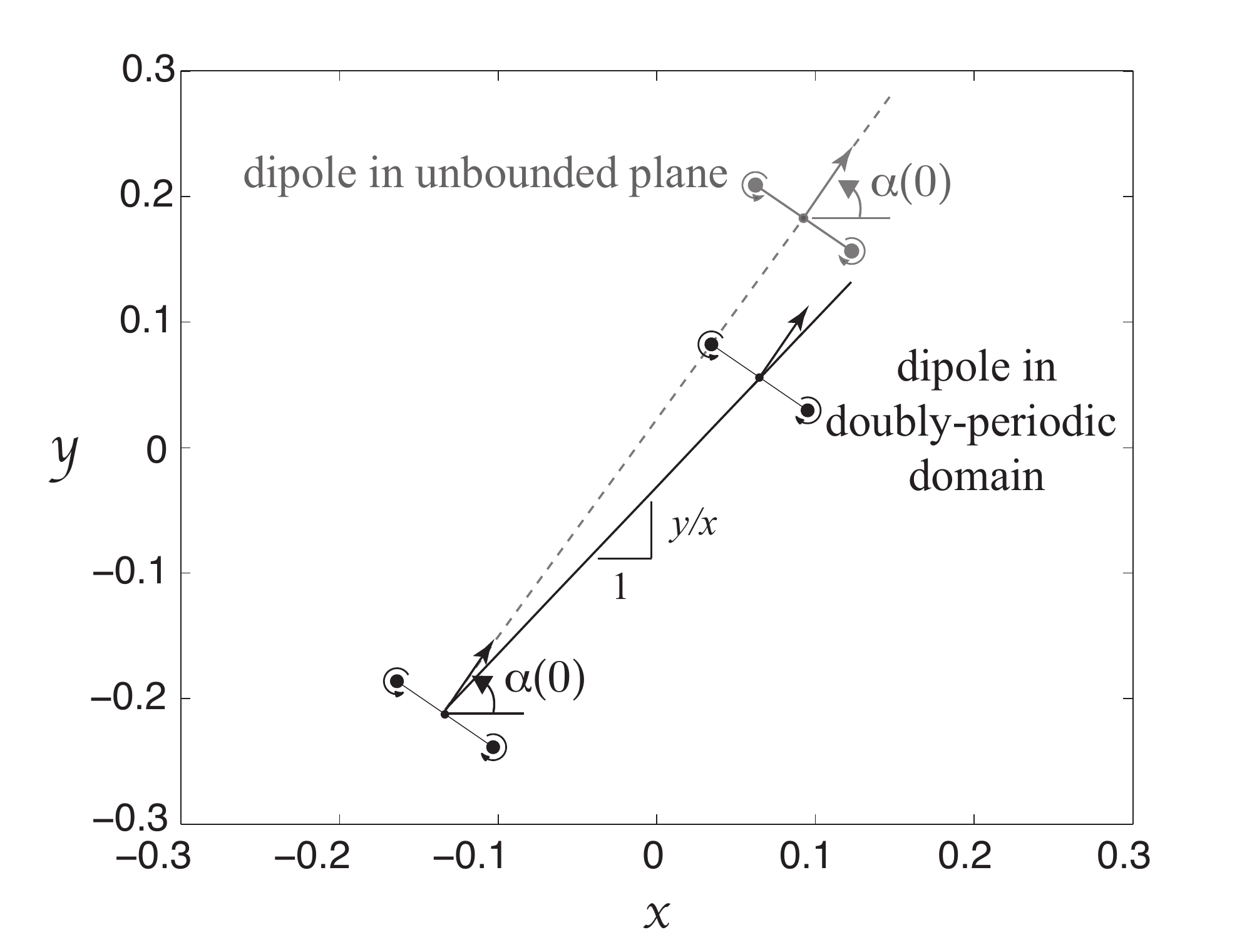}
\caption{\footnotesize Comparison between trajectories of a single dipole in unbounded plane and in doubly-periodic domain for $\omega_1=0.2$, $ \omega_2=0.2\ii$ and $\alpha(0)=\pi/3$. 
Black color denotes the path taken by the dipole in a doubly-periodic domain while grey color denotes the path taken by the dipole in unbounded plane. In both cases, the orientation $\alpha$ remains constant for all time but in the doubly-periodic domain, the slope of the trajectory traced by the dipole center is not equal to $\alpha$.}
  \label{fig:deviate}
  \end{center}
\end{figure}

We consider the seemingly simple case of a single dipole in a doubly-periodic domain. The dipole is only subject to its self-induced velocity and the velocity induced by its own images. That is to say, $w_{n,\mathrm o}$ and $\lambda_n$ are identically zero and equations \eqref{eq:formulation:eomZn}, \eqref{eq:formulation:eomAlphaN} take the form
\begin{equation}
\label{eq:formulation:eomOneDipole}
	\dot{\conj{z}}=\frac{\Gamma}{2\pi\ii}\left[\zeta\left(-\ii \ell \eee{\ii \alpha}\right)-\left(\frac{\pi\conj{\omega}_{1}}{\Delta\omega_{1}}-\frac{\eta_{1}}{\omega_{1}}\right)\ii \ell \eee{ \ii \alpha}-\frac{\pi}{\Delta}\ii \ell\eee{-\ii \alpha}\right], \qquad \dot{\alpha}=0.
\end{equation}
Note that the second term on the right-hand side of the first equation is identically zero in a square domain. This equation can be integrated in closed form,
\begin{equation}
\label{eq:SolOneDipolePos}
	\conj{z}=\frac{\Gamma}{2\pi\ii}\left[\zeta\left(-\ii \ell \eee{\ii \alpha}\right)-\left(\frac{\pi\conj{\omega}_{1}}{\Delta\omega_{1}}-\frac{\eta_{1}}{\omega_{1}}\right)\ii \ell \eee{ \ii \alpha}-\frac{\pi}{\Delta}\ii \ell \eee{-\ii \alpha}\right]t+\conj{z}(0), \qquad 	\alpha=\alpha(0).
\end{equation}
Here, $z(0)$ and ${\alpha}(0)$ are the initial position and orientation of the dipole respectively. That is to say, a single dipole always moves in a straight line with its orientation angle $\alpha$ unchanged. However, the slope of the linear trajectory of the dipole center is not in the direction of the orientation angle $\alpha$, see Figure~\ref{fig:deviate}, except for special initial conditions such as $\alpha(0)=k\pi/2$, $k$ arbitrary integer. The reason for this discrepancy between the slope of the dipole trajectory and its orientation is due to the periodic effect brought by the dipole images included in the $\zeta$-function.

\begin{figure}[!t]
	\begin{center}
	\subfigure[in doubly-periodic domain]{\includegraphics[scale=0.33]{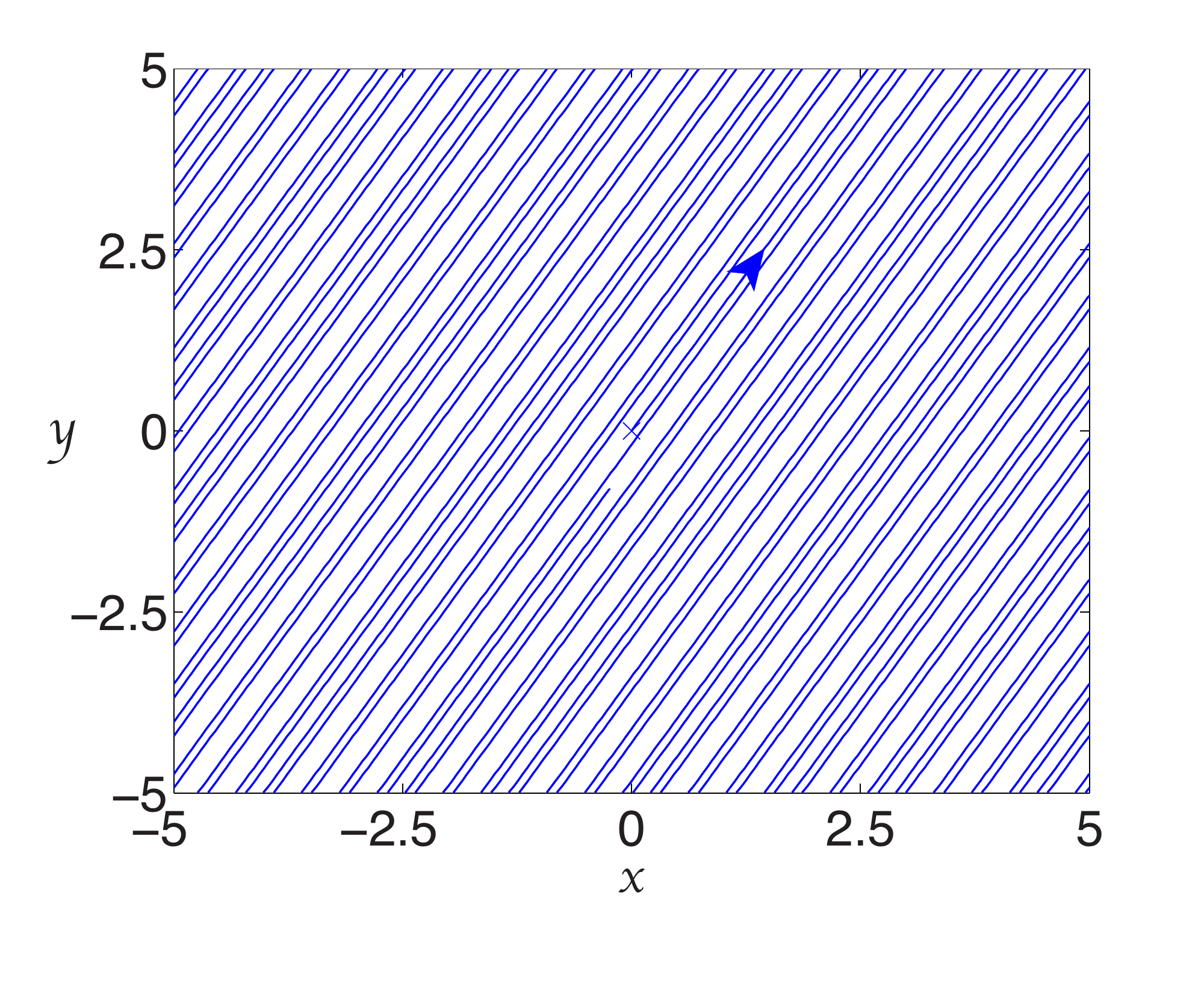}}\hspace{0.5in}
	\subfigure[on torus]{\includegraphics[scale=0.33]{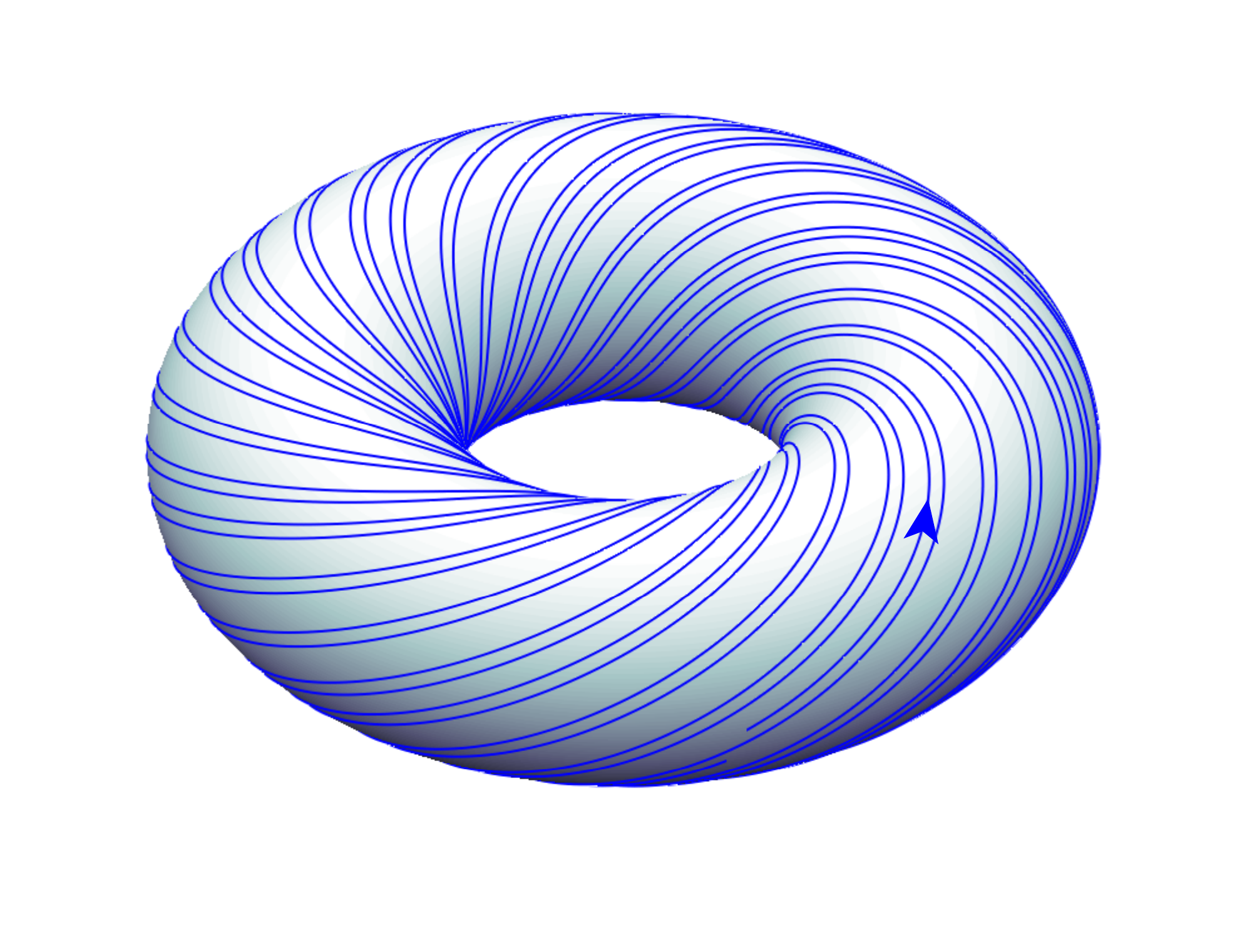}}
	\end{center}
\caption[]{Aperiodic trajectory that densely fills the whole domain:  $($a$)$ trajectory depicted in doubly-periodic domain $($b$)$ same trajectory depicted on a  torus.
Parameter values are: $\ell=1/2\pi$, $\omega = 5$, $z(0)=0$, $\alpha(0)=\pi/3$.} 
	\label{fig:onedipoleaperiodic}
\end{figure}

A single dipole affords two distinct types of dynamical behavior: aperiodic and periodic. For the former type, the single dipole traces out path which fills up the whole domain as depicted in Figure~\ref{fig:onedipoleaperiodic}(a). Here, the dipole never returns to the same location it visited thus the term aperiodic. Note that in this and the coming sections, we consider a doubly-periodic domain with half period $\omega_1=5$ and $\omega_2=5\ii$.  In all simulations, the strength $\Gamma$ of the dipoles' vortices is set to unity and the finite dipole length $\ell = 1/2\pi$. To emphasize the doubly-periodic nature of the domain, the dipole trajectory  is plotted on a torus as done in Figure~\ref{fig:onedipoleaperiodic}(b). The torus is obtained by applying the linear transformation $(x/\omega_1,y/\abs{\omega_2})\rightarrow(u,v)\in[-\pi,\pi]$ and $(\omega_1,\abs{\omega_2})\rightarrow(R,r)$, where $u,v$ are the angles and $R,r$ are the radii defining the torus. More specifically, a point $(X,Y,Z)$ on the torus is given by $X= (R+r\cos{v})\cos{u}$, $Y= (R+r\cos{v})\sin{u}$ and $Z=r\sin{v}$. The ratio of $r$ to $R$ is selected to be 1 to 2 for clarity of exposition.

\begin{figure}[!t]
	\begin{center}
	\subfigure[Rational ratios of $q/p$ give $\alpha(0)$ that produce periodic trajectories]{\includegraphics[scale=0.5]{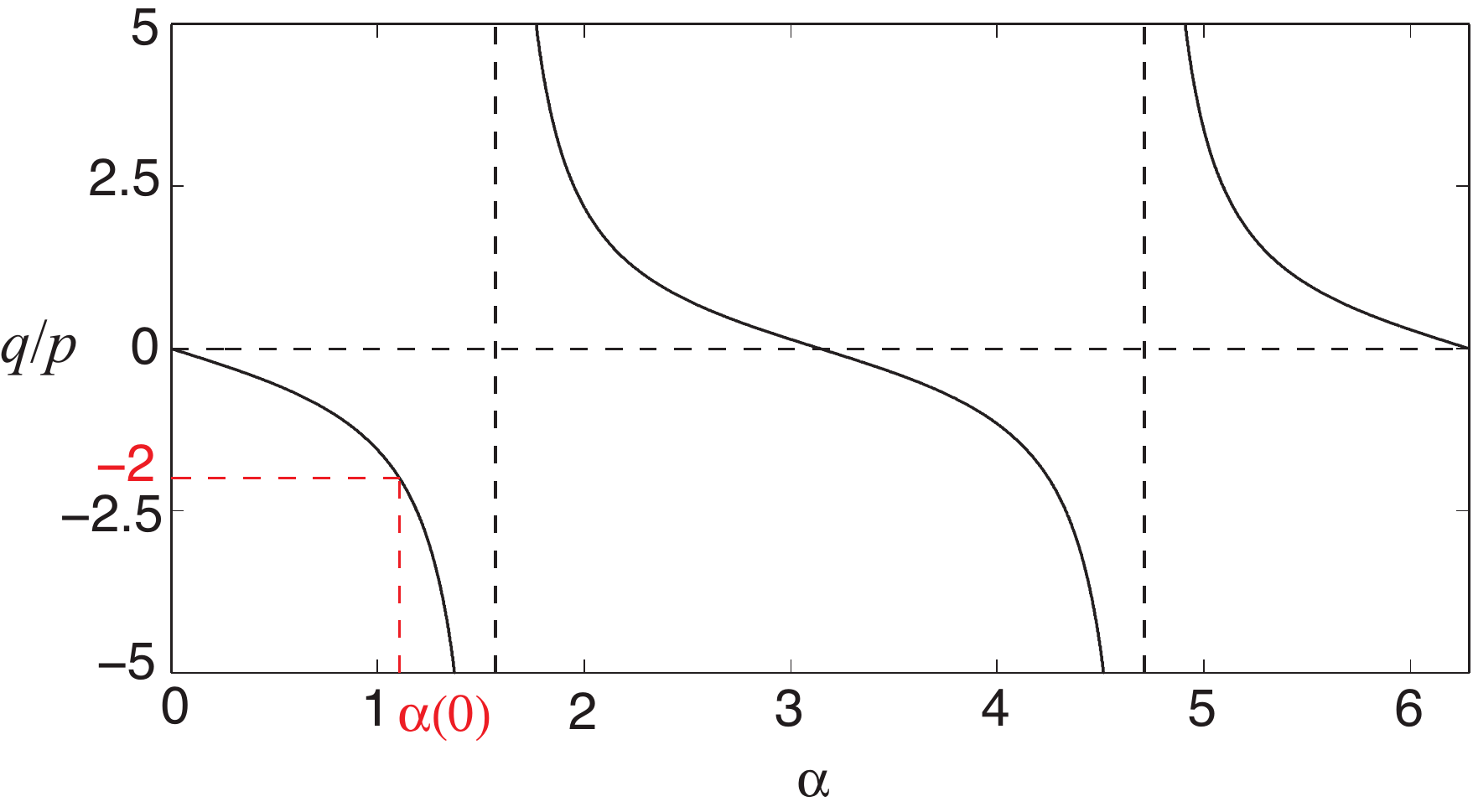}}\\
	\subfigure[in doubly-periodic domain]{\includegraphics[scale=0.33]{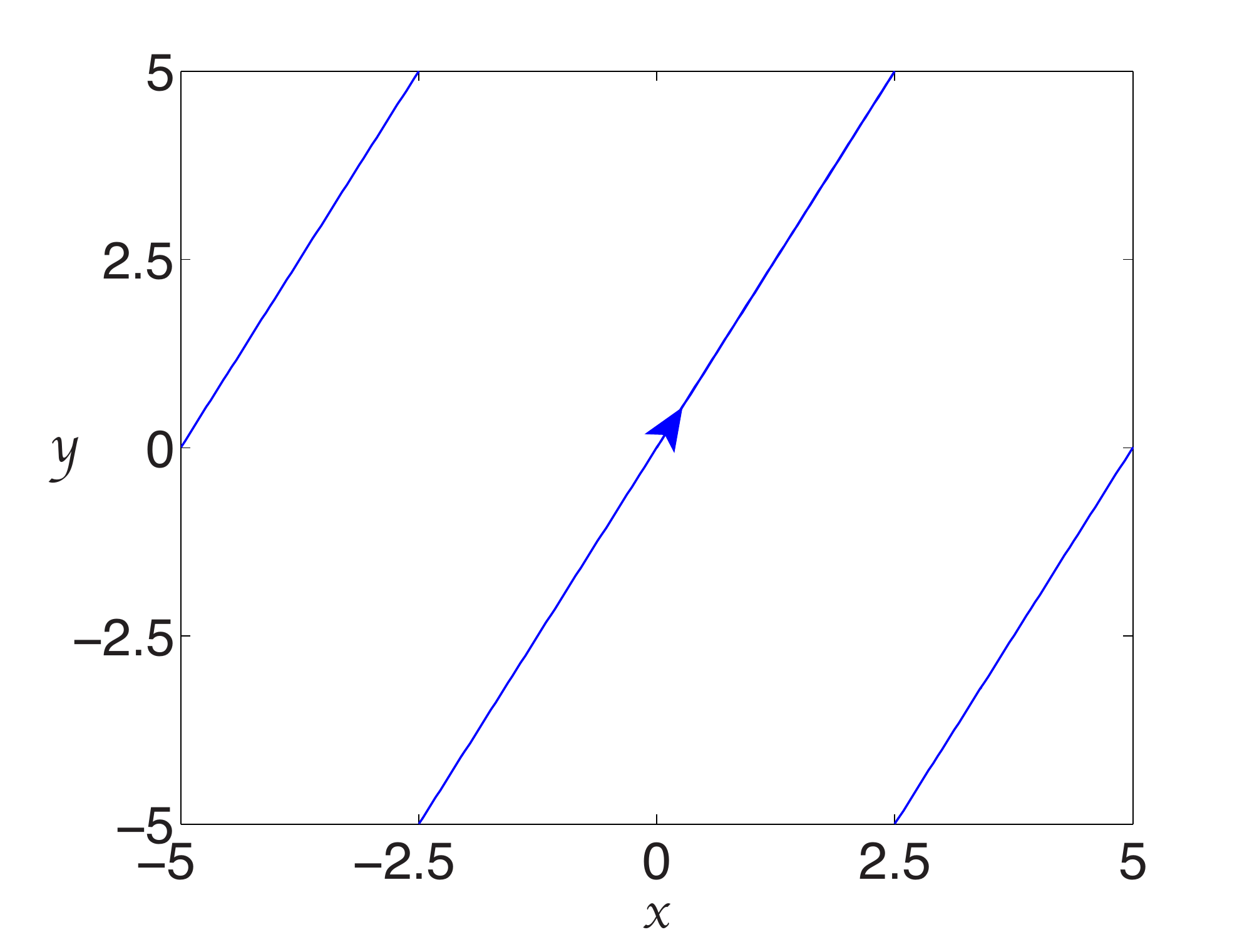}}\hspace{0.5in}
	\subfigure[on torus]{\includegraphics[scale=0.33]{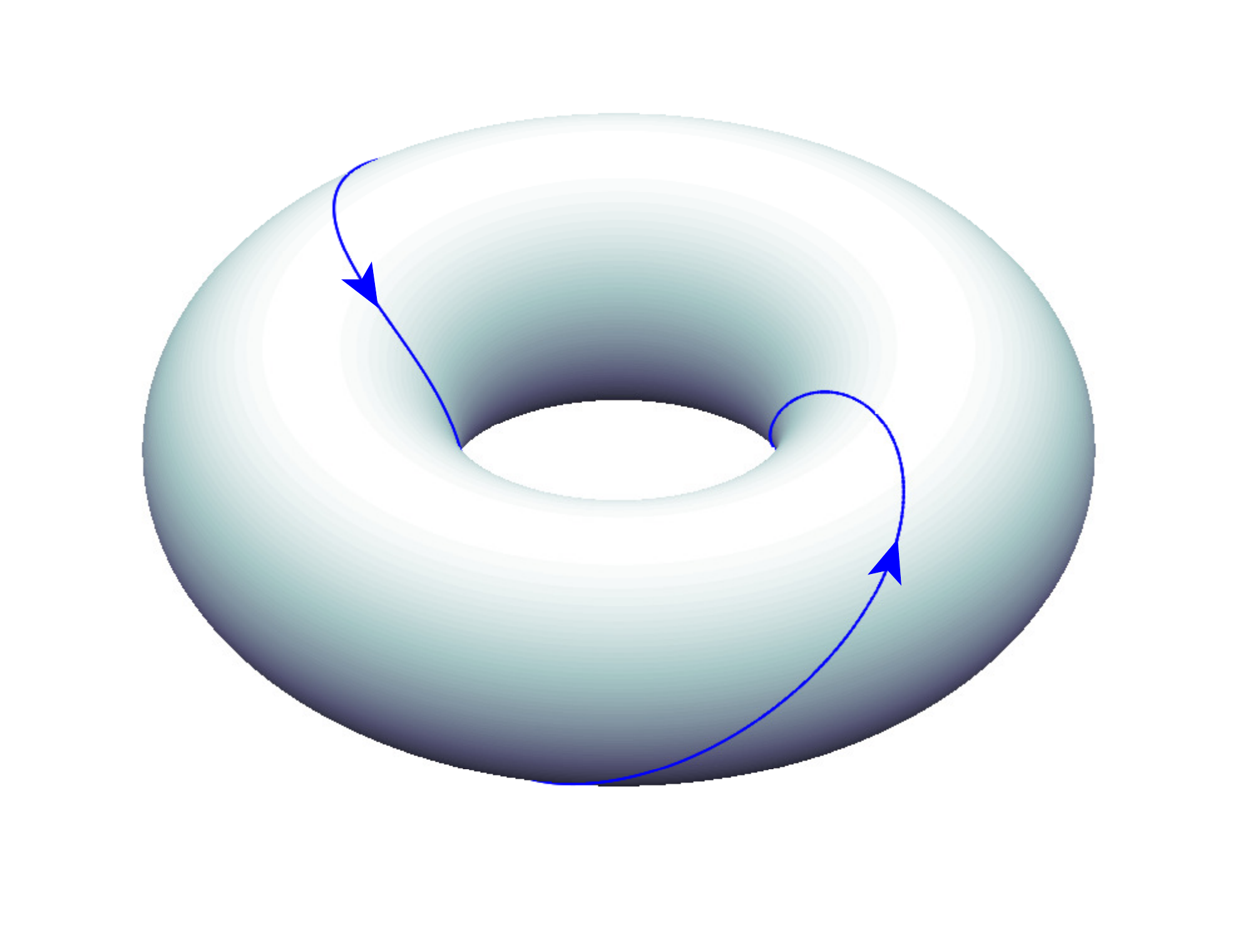}}
\end{center}
\caption[]{ Periodic trajectory of a single dipole in doubly-periodic domain: $($a$)$ according to equation~\eqref{eq:PeriodicSol2}, rational values of $q/p$ give $\alpha(0)$ that produce periodic trajectories.
$($b$)$ periodic trajectory for $p/q = -2$ and parameter values  $\ell=1/2\pi$ and $\omega=5$. The value of $\alpha(0)$ is obtained from plot $($a$)$. $($b$)$ same trajectory depicted on a torus. 
} 
	\label{fig:onedipoleperiodic}
\end{figure}

The aperiodic behavior in Figure~\ref{fig:onedipoleaperiodic} seems to be the generic behavior for arbitrary initial conditions. We then ask for what initial conditions (if any), the single dipole exhibits periodic solutions.
That is to say, we look for solutions that satisfy the condition
\begin{equation}
\label{eq:PeriodicSol}
	\conj{z}(T)=\conj{z}(0)+2p\omega_1+2q\omega_2,
\end{equation}
where $p$ and $q$ are integers and $T$ is the period of the motion. Using~\eqref{eq:SolOneDipolePos}, the above equation amounts to
\begin{equation}
\label{eq:PeriodicSol2}
	\frac{\Gamma}{2\pi\ii}\left[\zeta\left(-\ii \ell \eee{\ii \alpha}\right)-\left(\frac{\pi\conj{\omega}_{1}}{\Delta\omega_{1}}-\frac{\eta_{1}}{\omega_{1}}\right)\ii \ell \eee{ \ii \alpha}-\frac{\pi}{\Delta}\ii \ell\eee{-\ii \alpha}\right]T
	=2p\omega_1+2q\omega_2, \qquad p, q \in \mathbb{Z}.
\end{equation}
In a square domain, $\omega_1 = -\ii\omega_2 = \omega$,
\eqref{eq:PeriodicSol2} is satisfied only when the imaginary and real part on the left hand side of the equation is in rational ratio of $q$ to $p$. The corresponding value of $\alpha$ can be evaluated numerically for different $q$ to $p$ ratio. See Figure~\ref{fig:onedipoleperiodic} for a depiction of a periodic trajectory for $q/p=-2$. 

We close this section by noting that in the limiting case of an infinitely large domain, $\omega_1$, $\omega_2$ $\rightarrow$ $\infty$, $\zeta(z)$ reduces to $1/z$ and $\Delta$$\rightarrow$ 0 in \eqref{eq:PeriodicSol}. Left hand side of \eqref{eq:PeriodicSol} thus becomes
\begin{equation}
\label{eq:LimitGeodesic}
	\lim_{\omega_1,\omega_2  \rightarrow \infty}\frac{\Gamma}{2\pi\ii}\left[\zeta\left(-\ii \ell \eee{\ii \alpha}\right)-\left(\frac{\pi\conj{\omega}_{1}}{\Delta\omega_{1}}-\frac{\eta_{1}}{\omega_{1}}\right)\ii \ell \eee{ \ii \alpha}-\frac{\pi}{\Delta}\ii \ell\eee{-\ii \alpha}\right]T=\frac{\Gamma \eee{-\ii \alpha}}{2 \pi \ell}T.
\end{equation}
The ratio of the imaginary and real part of the above expression is -$\tan{\alpha}$ and the self-induced velocity reduces to the case of an unbound plane, represented by \eqref{eq:formulation:wnsi}.


\section{Collision, no-collision and synchronization of two dipoles}
\label{sec:interactions}

In this section, we consider the interaction of two finite dipoles ($z_1$, $\alpha_1$) and ($z_2$,  $\alpha_2$) of equal length $\ell_1=\ell_2=\ell$ and equal strength $\Gamma_1=\Gamma_2=\Gamma$, we describe three distinct dynamical behavior: collision, collision-avoidance and motion synchronization of the two dipoles. These nontrivial interactions arise solely from hydrodynamic coupling.

For concreteness, we write the equations of motion for the two dipole system by substituting \eqref{eq:formulation:wnsperiodic} and \eqref{eq:formulation:wnoperiodic}  into~\eqref{eq:formulation:eomZn} and \eqref{eq:formulation:eomAlphaN}. To this end, one has
\begin{equation}
\label{eq:formulation:eomTwoDipoles}
\begin{split}
\dot{\conj{z}}_1 & = \dfrac{\Gamma}{2\pi\ii}\Bigl\{\zeta(-\ii \ell \eee{\ii \alpha_{1}})
+ \dfrac{1}{2}\Bigl[ \zeta(z_{1,\textrm{l}}-z_{2,\textrm{l}}) 
						- \zeta(z_{1,\textrm{l}}-z_{2,\textrm{r}}) 
						+ \zeta(z_{1,\textrm{r}}-z_{2,\textrm{l}}) 
						- \zeta(z_{1,\textrm{r}}-z_{2,\textrm{r}}) \Bigr] \\[2ex]
& \hspace{0.5in} - \ii \ell \left(\eee{\ii\alpha_1} +  \eee{\ii\alpha_2}\right)\left(\frac{\pi\conj{\omega_{1}}}{\Delta\omega_{1}}-\frac{\eta_{1}}{\omega_{1}} \right)	
			 -\ii \ell \left( \eee{-\ii\alpha_1} + \eee{-\ii\alpha_2}\right)\dfrac{\pi}{\Delta}\Bigr\} ,
\end{split}
\end{equation}
and
\begin{equation}
\label{eq:formulation:eomTwoDipolesAngle}
\begin{split}
	  \dot{\alpha}_1 & =\text{Re} \Bigl\{\frac{\Gamma}{4\pi\ii}\eee{\ii \alpha_{1}}\Bigl[ \zeta(z_{1,\textrm{r}}-z_{2,\textrm{l}}) 
						- \zeta(z_{1,\textrm{r}}-z_{2,\textrm{r}}) 
						- \zeta(z_{1,\textrm{l}}-z_{2,\textrm{l}}) 
						+ \zeta(z_{1,\textrm{l}}-z_{2,\textrm{r}}) \Bigr]\Bigr\}.
\end{split}
\end{equation}
Similar equations hold for $z_2$ and $\alpha_2$. These equations form a system of six coupled nonlinear ordinary differential equations which we solve numerically using a standard Runge-Kutta solver with variable time step. For a fixed set of parameter values $\ell_1 = \ell_2 = 1/2\pi$, $\omega_1 = \abs{\omega_2} = 5$, we vary the initial conditions $z_1(0), \alpha_1(0), z_2(0)$ and $\alpha_2(0)$. Depending on the choice of initial conditions, one obtains different dynamical behavior.  Note that the periodicity of the domain enriches the dynamics of the two dipoles and enables them to interact many times: when one dipole leaves the domain, it re-enters on the opposite side and continues to interact with the other dipole. In this sense, the two dipoles cannot diverge but do exhibit 
a range of dynamical behavior as summarized below.

\begin{figure}[!th]
\begin{center}
	\includegraphics[scale=0.33]{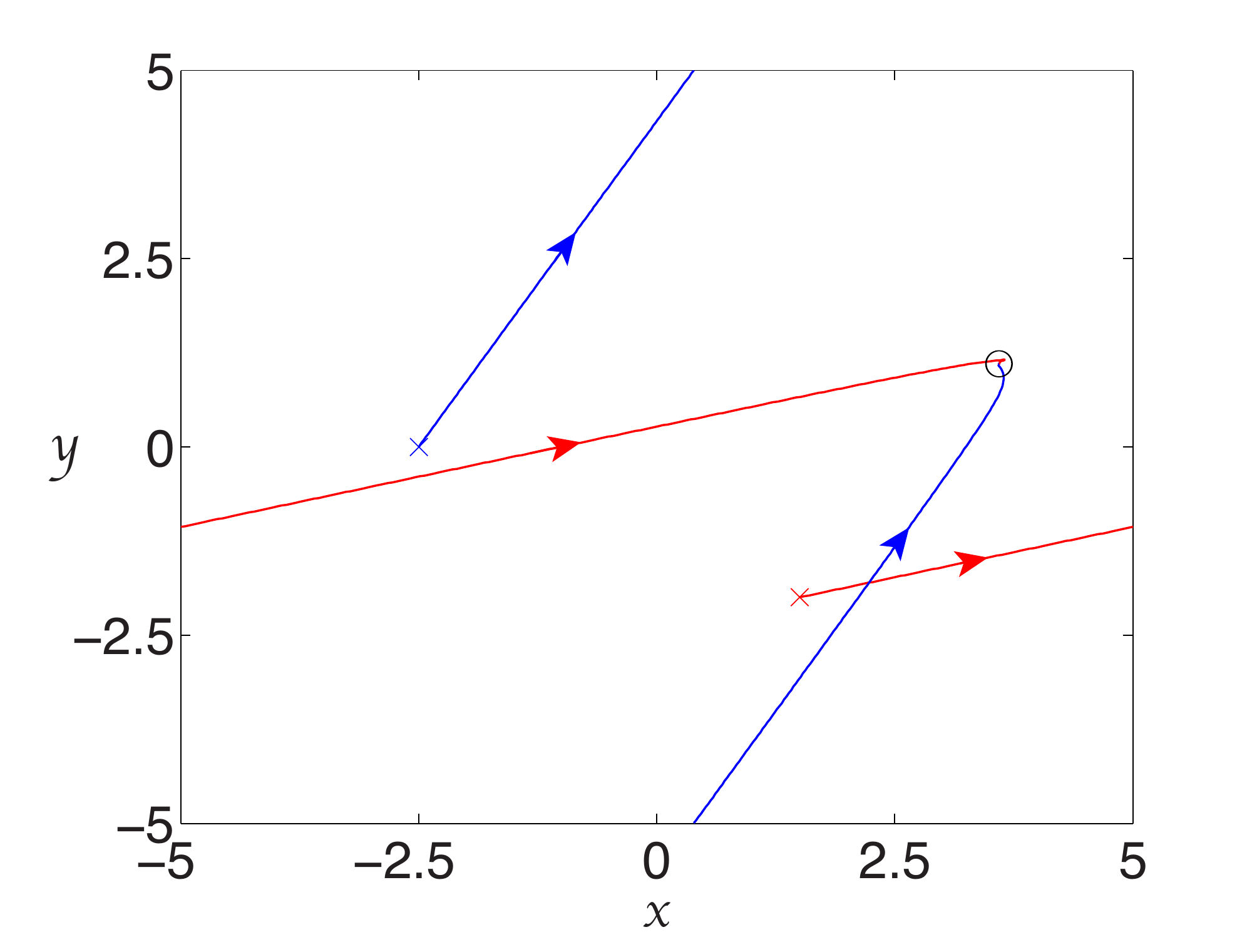}
\end{center}
\caption{ Collision of two dipoles in a doubly-periodic domain. The blue and red line represents the trajectories formed by the two dipoles. Parameter values are: $\ell=1/2\pi$, $z_1(0)=-2.5$, $z_2(0)=1.5-2\ii$, $\alpha_1(0)=\pi/3$, $\alpha_2(0)=\pi/12$.}
	\label{fig:2dipolescollision}
\end{figure}

\begin{figure}[!t]
\begin{center}
	\subfigure[in doubly-periodic domain]{ \includegraphics[scale=0.33]{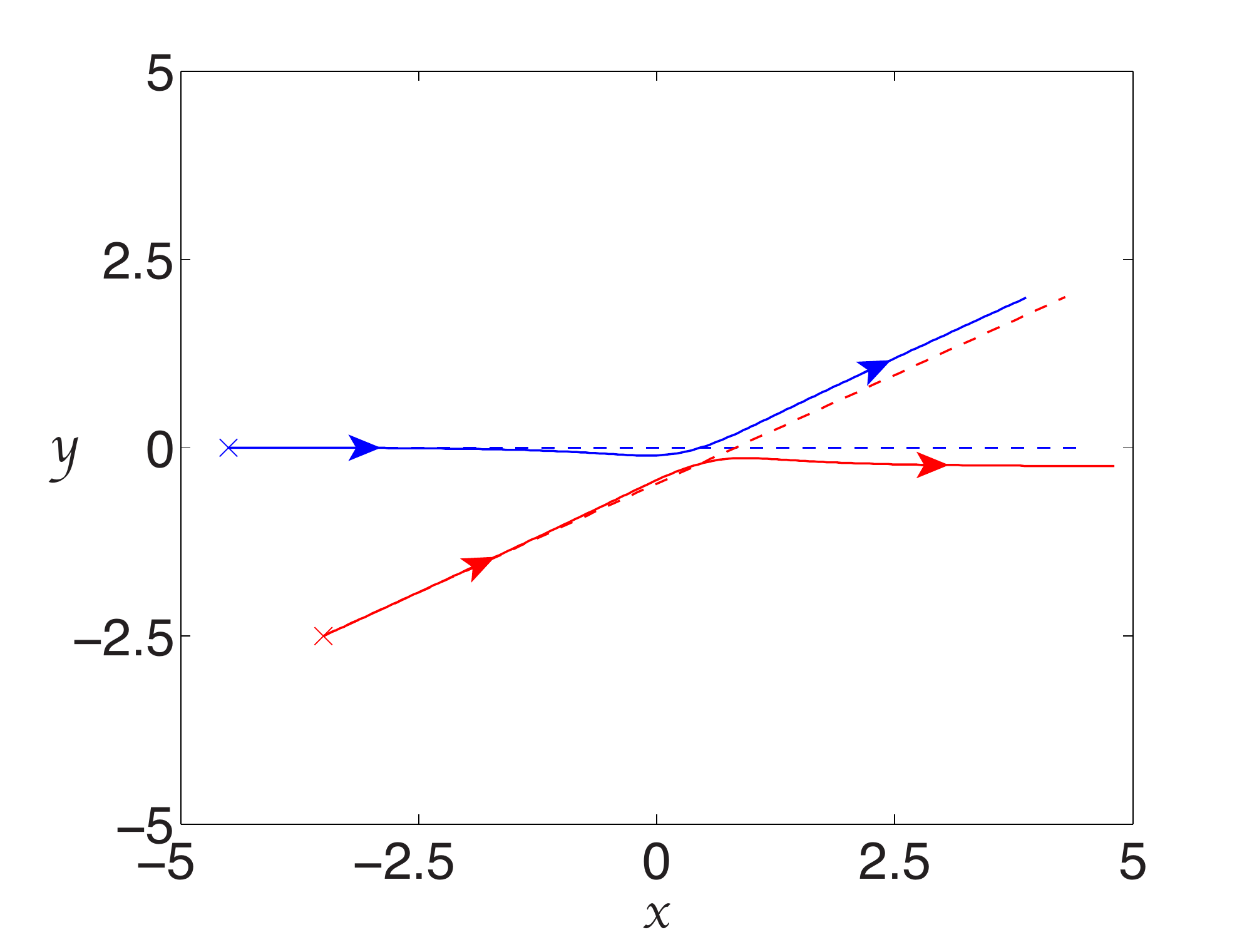}}	\hspace{0.5in}
	\subfigure[change in angles of orientation with time]{\includegraphics[scale=0.33]{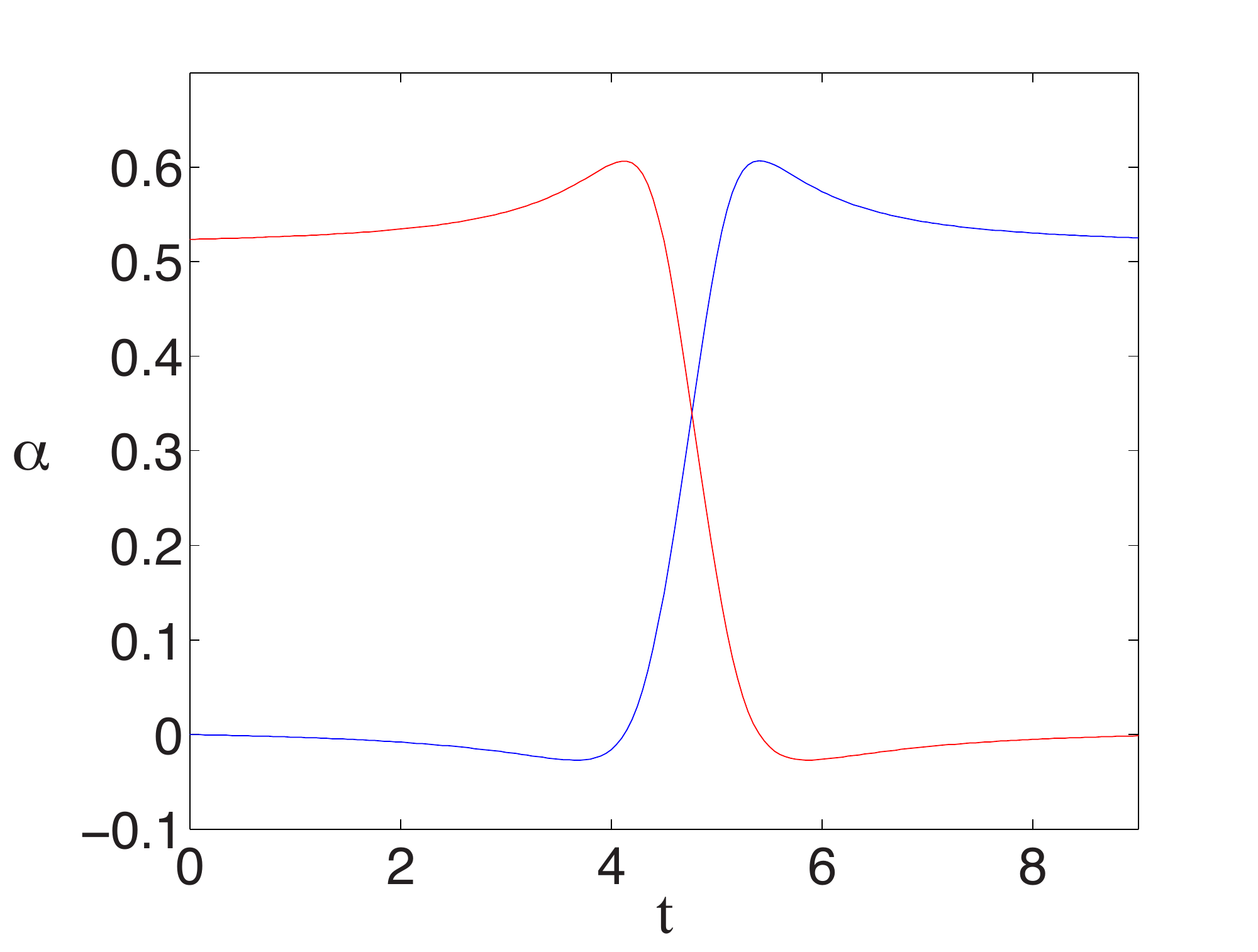}}
	 \subfigure[on torus]{\includegraphics[scale=0.33]{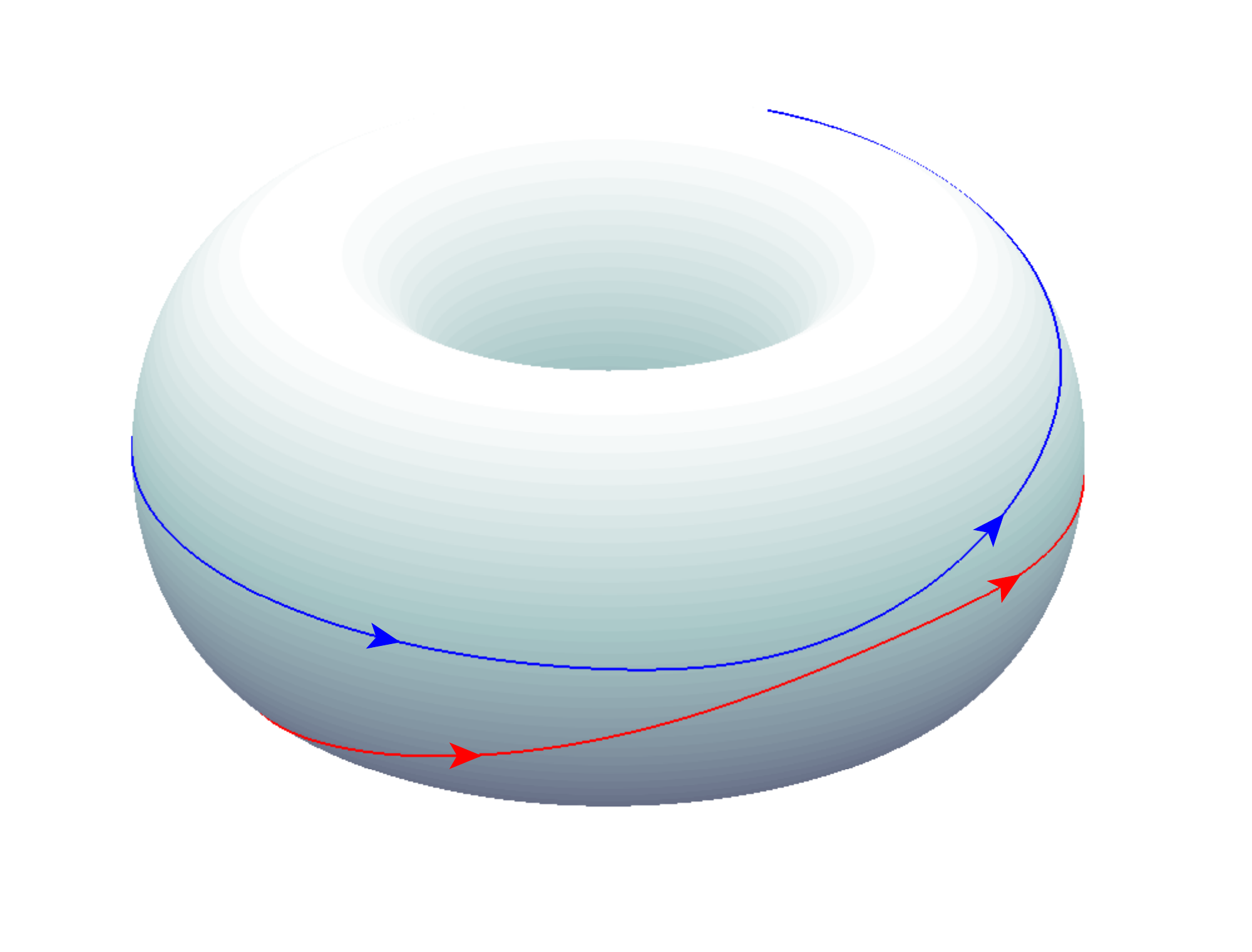}}	 
\end{center}
  	\caption[]{Collision-avoidance of two dipoles. $($a$)$  trajectories of the two dipoles. The blue and red dashed line are the paths of the dipoles if they act independently without interacting with each other.  $($b$)$ Orientation angles versus time. $($c$)$ trajectories depicted on a torus. Parameter values are: $\ell=1/2\pi$, $z_1(0)=-4.5$, $z_2(0)=-3.5-2.5\ii$, $\alpha_1(0)=0$, $\alpha_2(0)=\pi/6$. }
	\label{fig:2dipolesswitching}
\end{figure}
 	
\begin{figure}[tp]
\begin{center}
 	\subfigure[in doubly-periodic domain]{\includegraphics[scale=0.33]{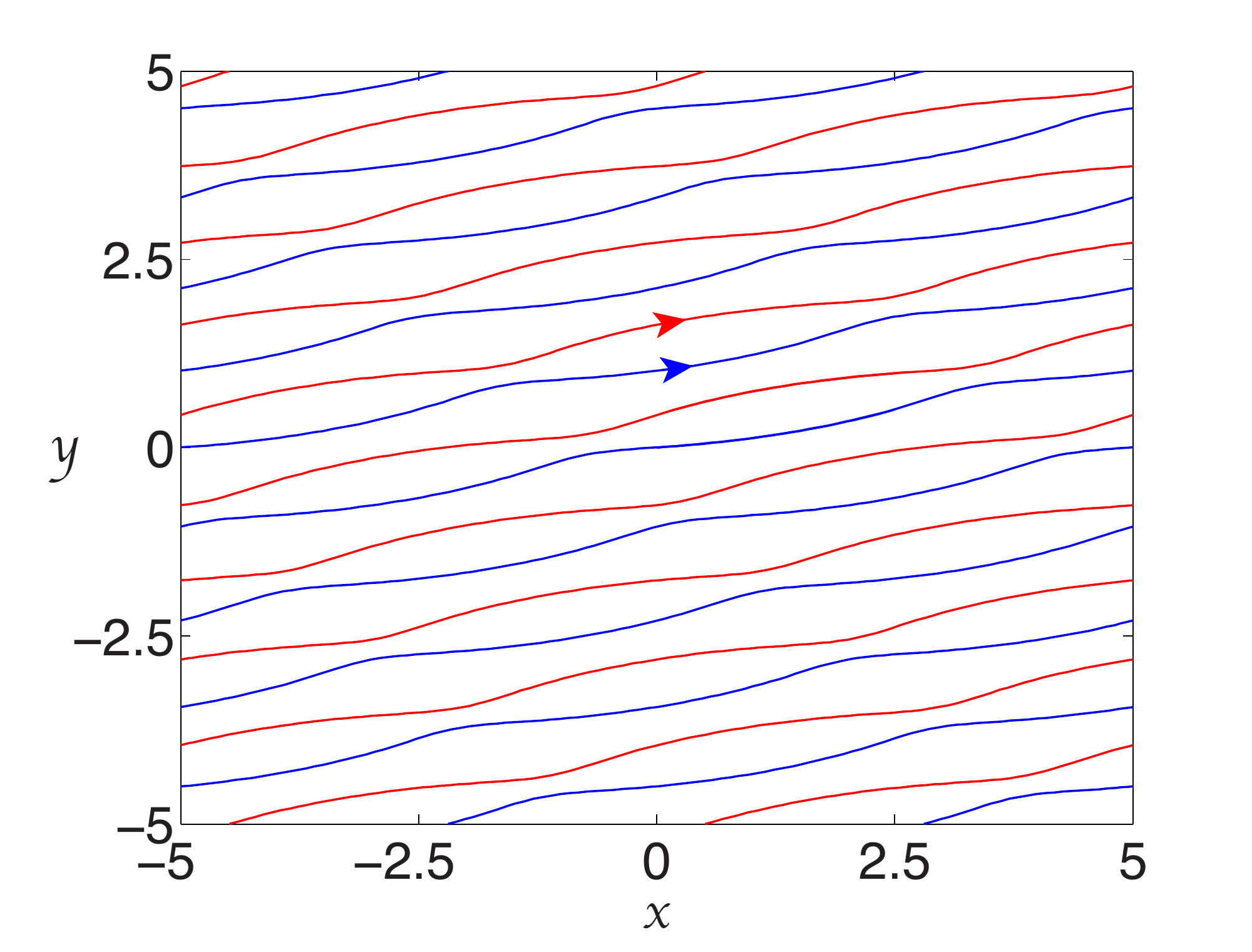}} \hspace{0.5in}
	\subfigure[on torus]{\includegraphics[scale=0.33]{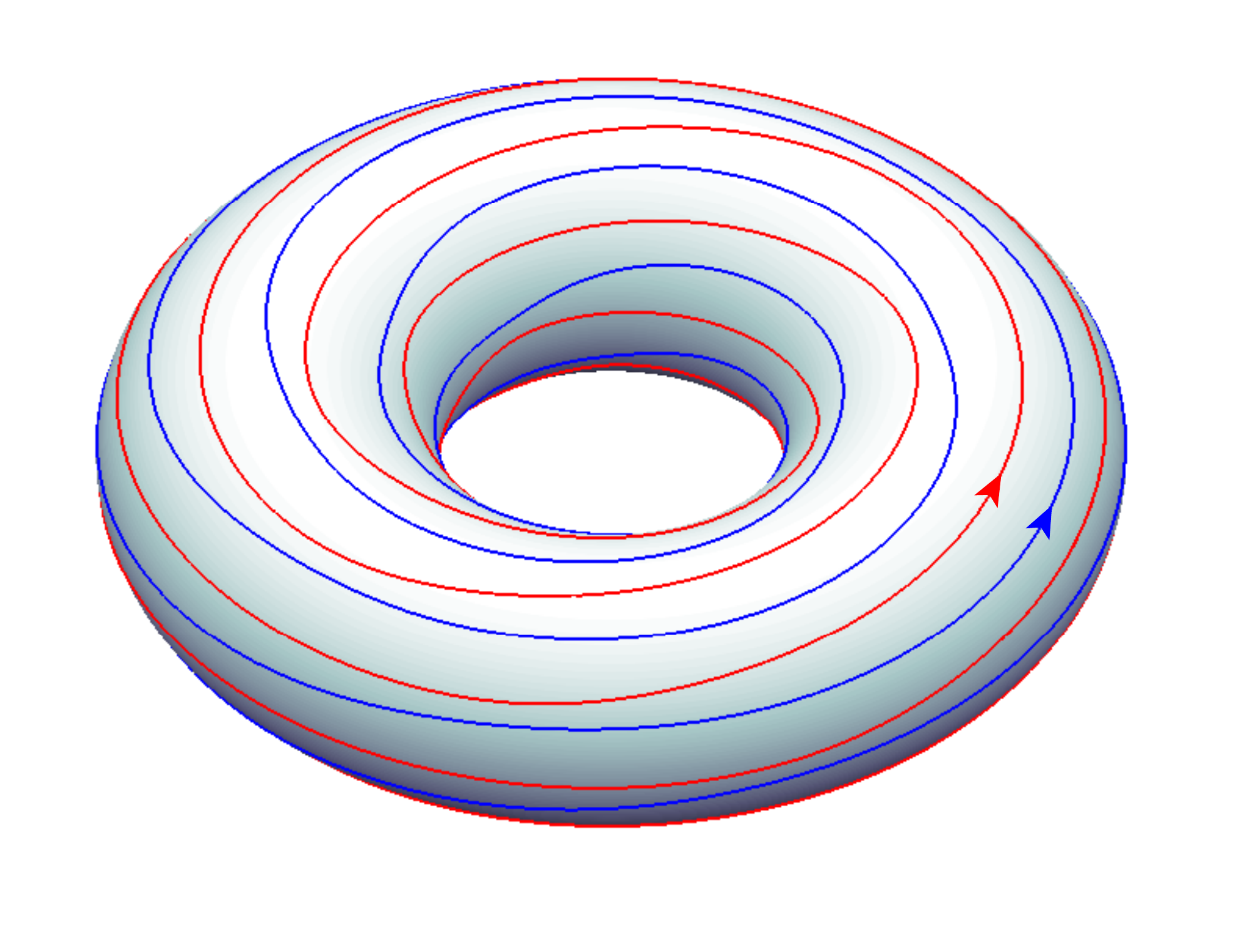}}
\end{center}
  	\caption[]{$($a$)$ Periodic trajectories of two dipoles. $z_1(0)=0$, $z_2(0)=0.1501+0.4901\ii$, $\alpha_1(0)=0$, $\alpha_2(0)=\pi/12$. $($b$)$ Same synchronization trajectories depicted on a torus.}
	\label{fig:2dipolessynchronization}
\end{figure}

\paragraph{Collision.} Two finite dipoles in doubly-periodic domain may collide in finite time.
When the dipoles collide, they form a fixed quadrupole.  
A typical example of dipole collision in a periodic domain is shown in Figure~\ref{fig:2dipolescollision}a.
A similar behavior is reported in \cite{Tchieu:prsa2012a} for two finite dipoles in an unbounded plane. However, the interactions of two dipoles  in an unbounded plane is simpler in the sense that one can identify the set of initial conditions that give rise to collision. In the doubly-periodic domain, the set of initial conditions that lead to collision seems to be dense in the space of all initial conditions (based on a range of numerical simulations not shown here for brevity).

\paragraph{Collision avoidance.} The hydrodynamic coupling between two dipoles could induce collision avoidance as shown in Figure~\ref{fig:2dipolesswitching}a. Figure~\ref{fig:2dipolesswitching}b plots the change in the dipoles' orientation as a function of time. As the two dipoles approach each other, their orientations change drastically and collision is avoided. The trajectories of the dipoles are represented on a torus in Figure~\ref{fig:2dipolesswitching}c. It is important to emphasize that this collision avoidance behavior is a result of the hydrodynamic coupling only with no external control. That is to say, the fluid medium plays the role of a collision avoidance mechanism
for certain approach conditions. Again, numerical evidence (results not shown here) suggests that  the set of initial conditions leading to collision avoidance is dense in the space of all initial conditions.

\begin{figure}[tp]
\begin{center}
	\subfigure[]{\includegraphics[scale=0.33]{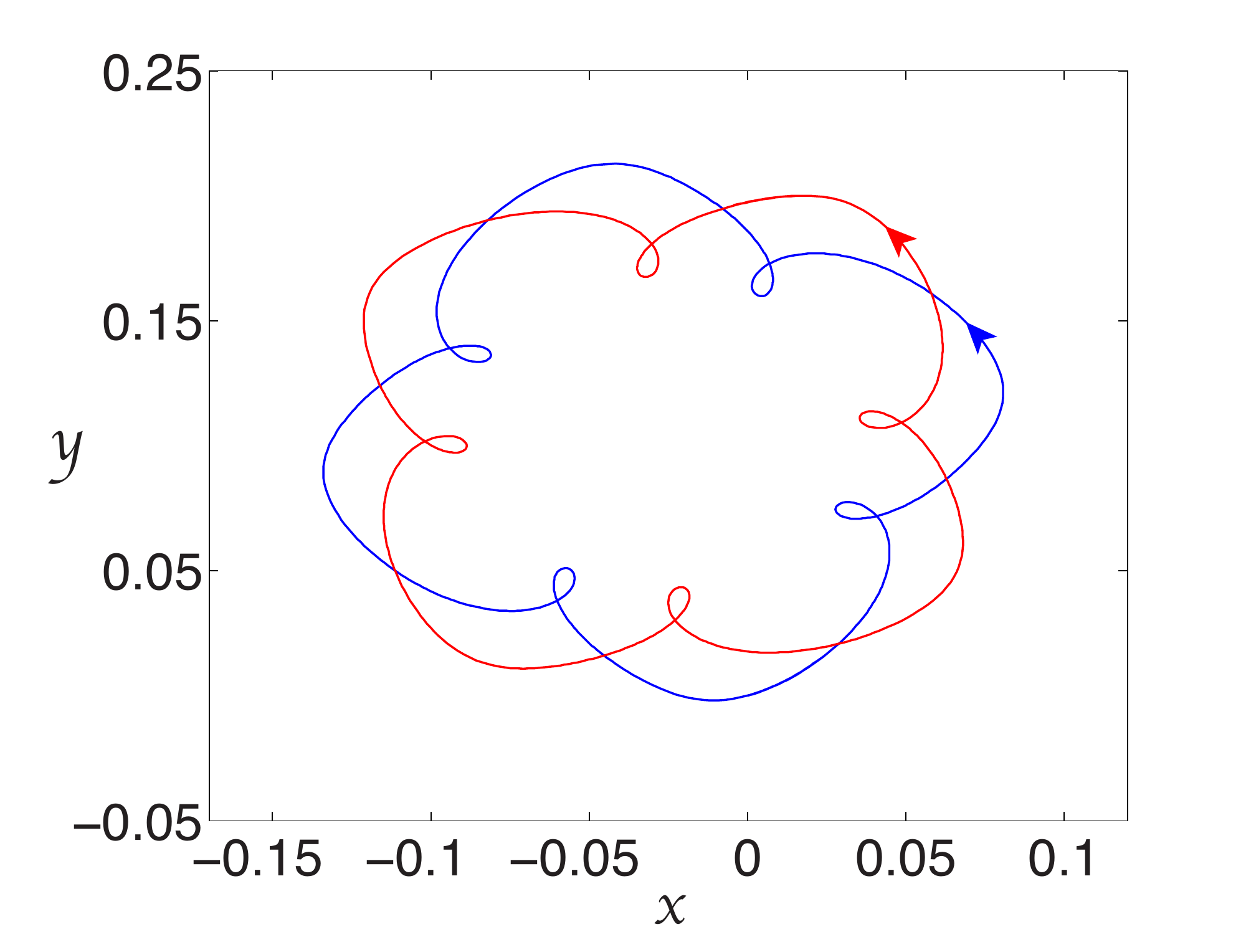}}
	\subfigure[]{\includegraphics[scale=0.33]{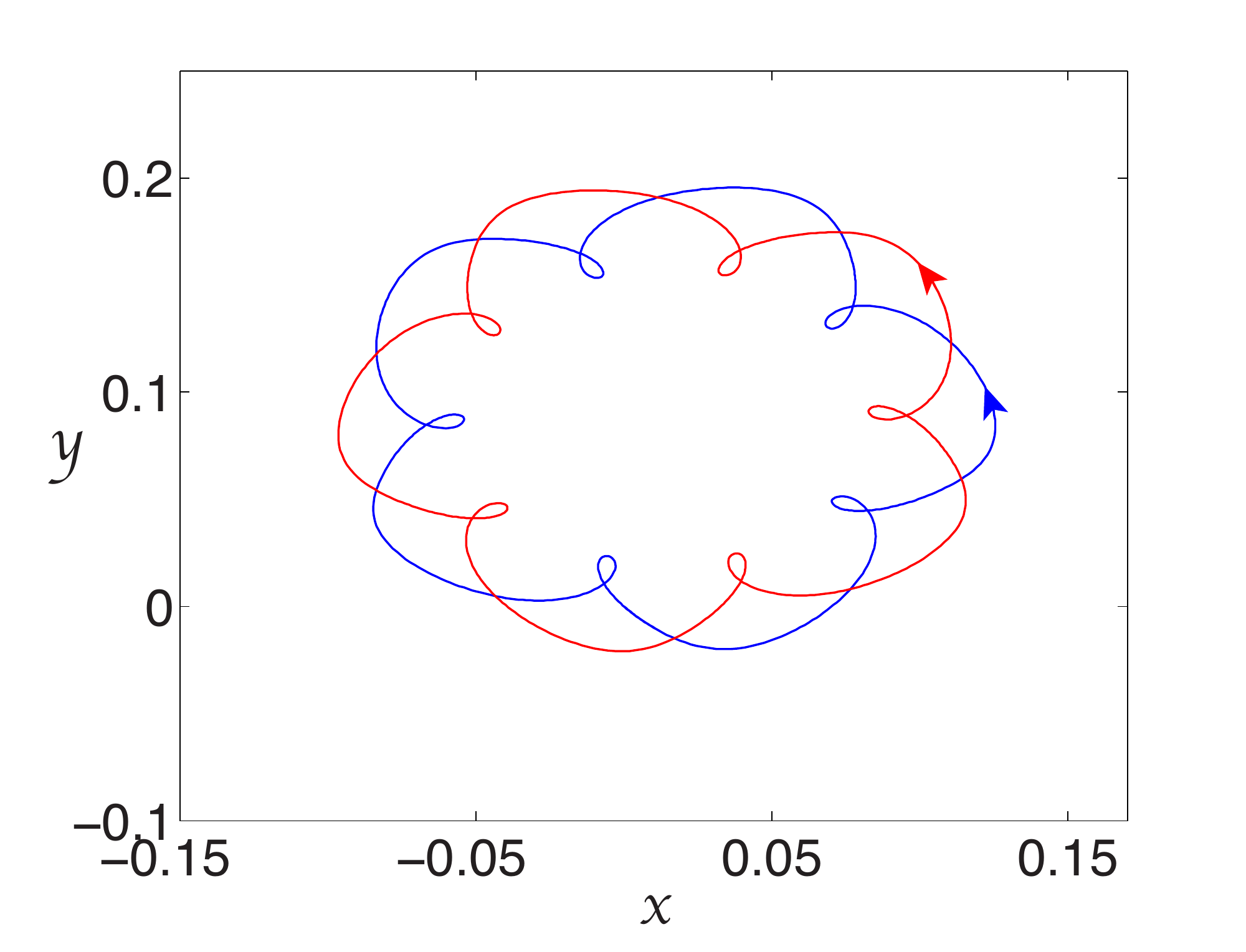}}
	\subfigure[]{\includegraphics[scale=0.33]{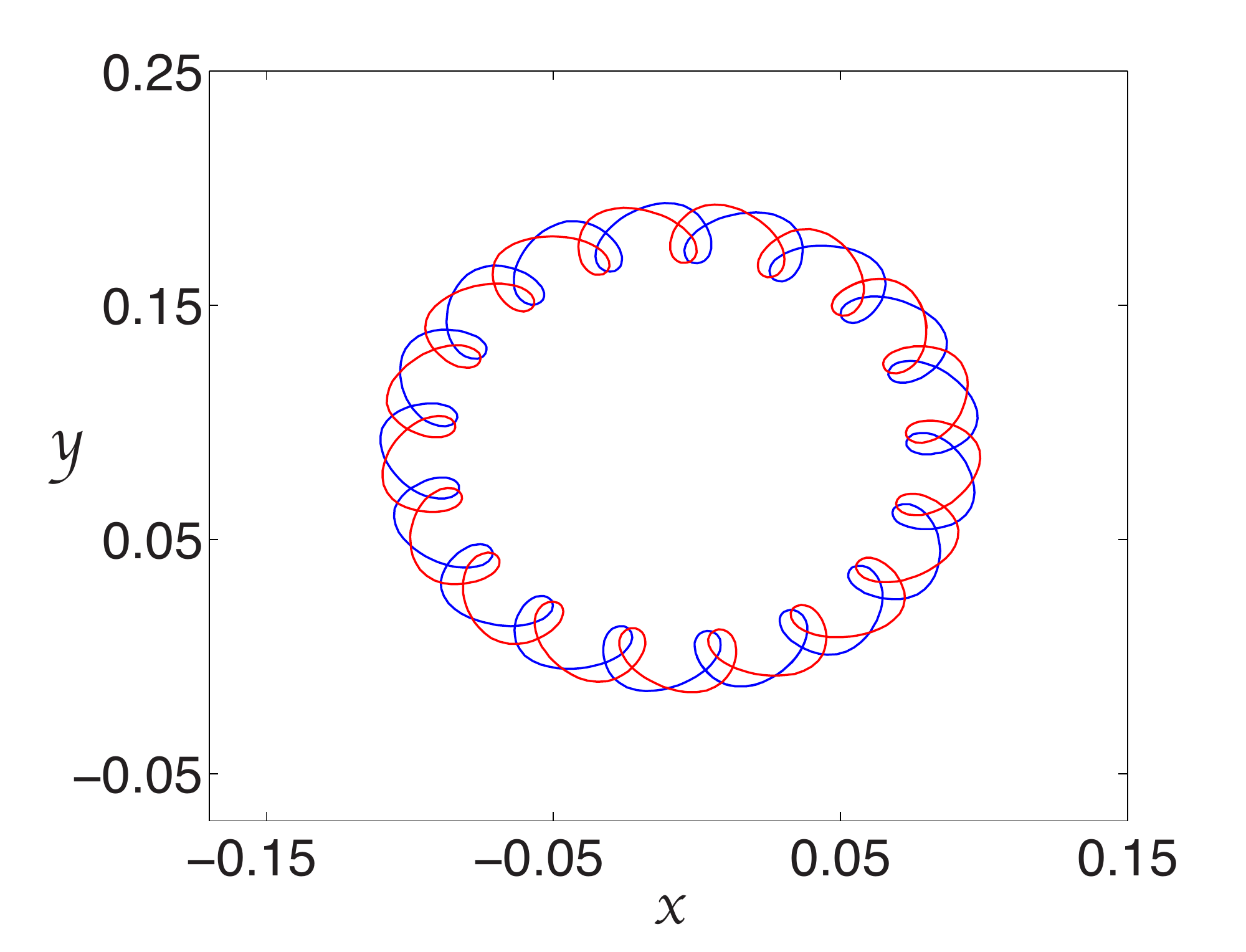}}
	\subfigure[]{\includegraphics[scale=0.33]{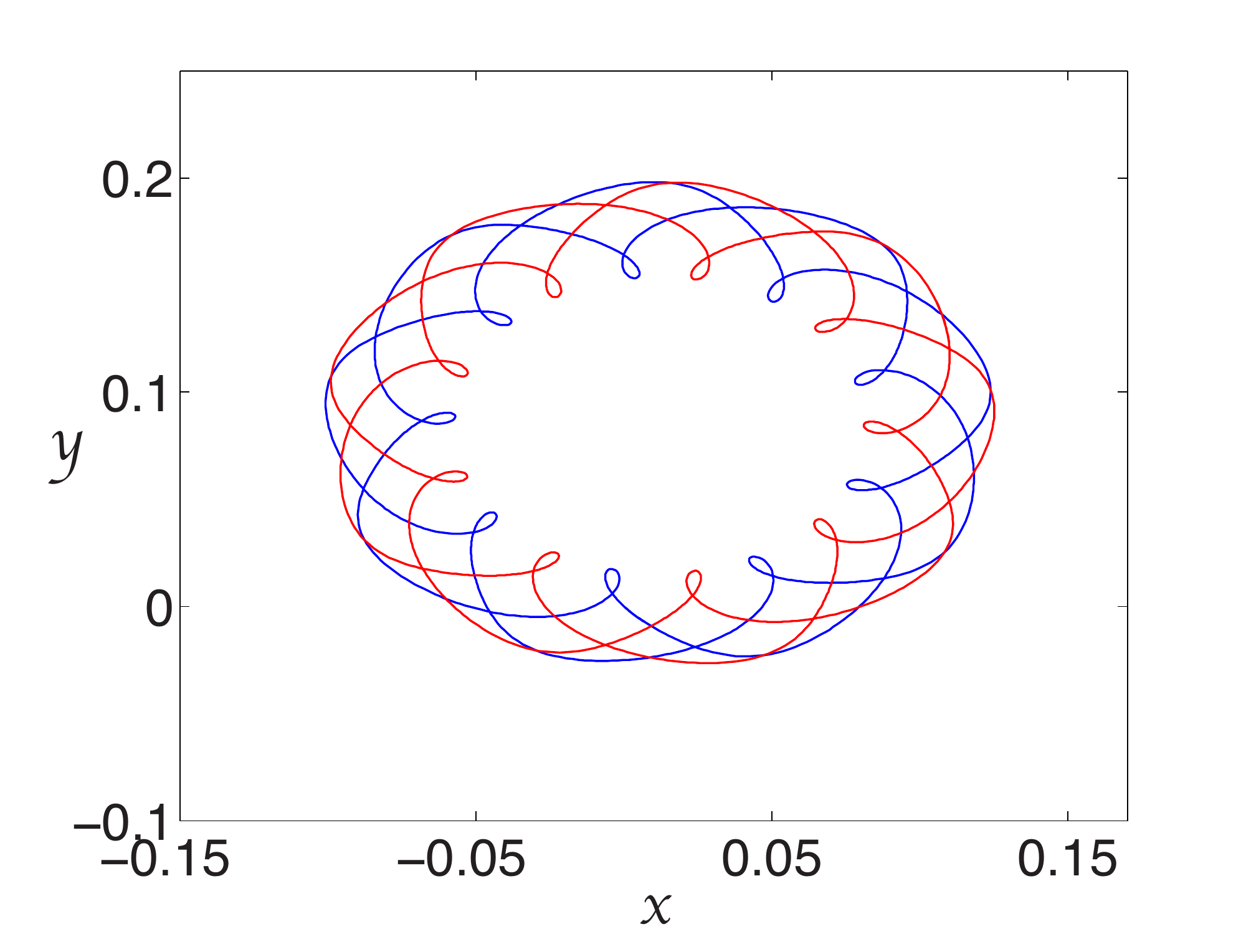}}
	\subfigure[]{\includegraphics[scale=0.33]{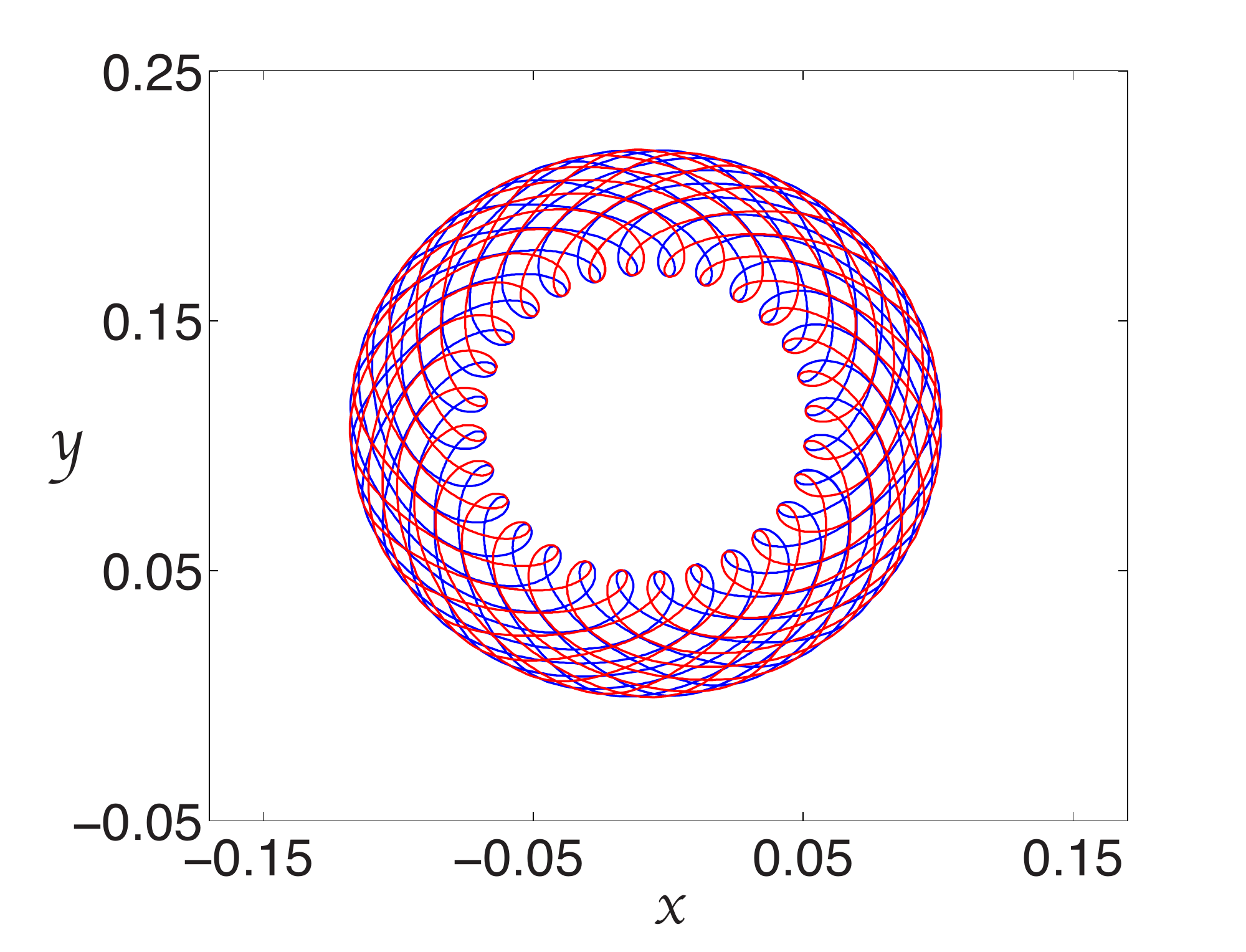}}
	\subfigure[]{\includegraphics[scale=0.33]{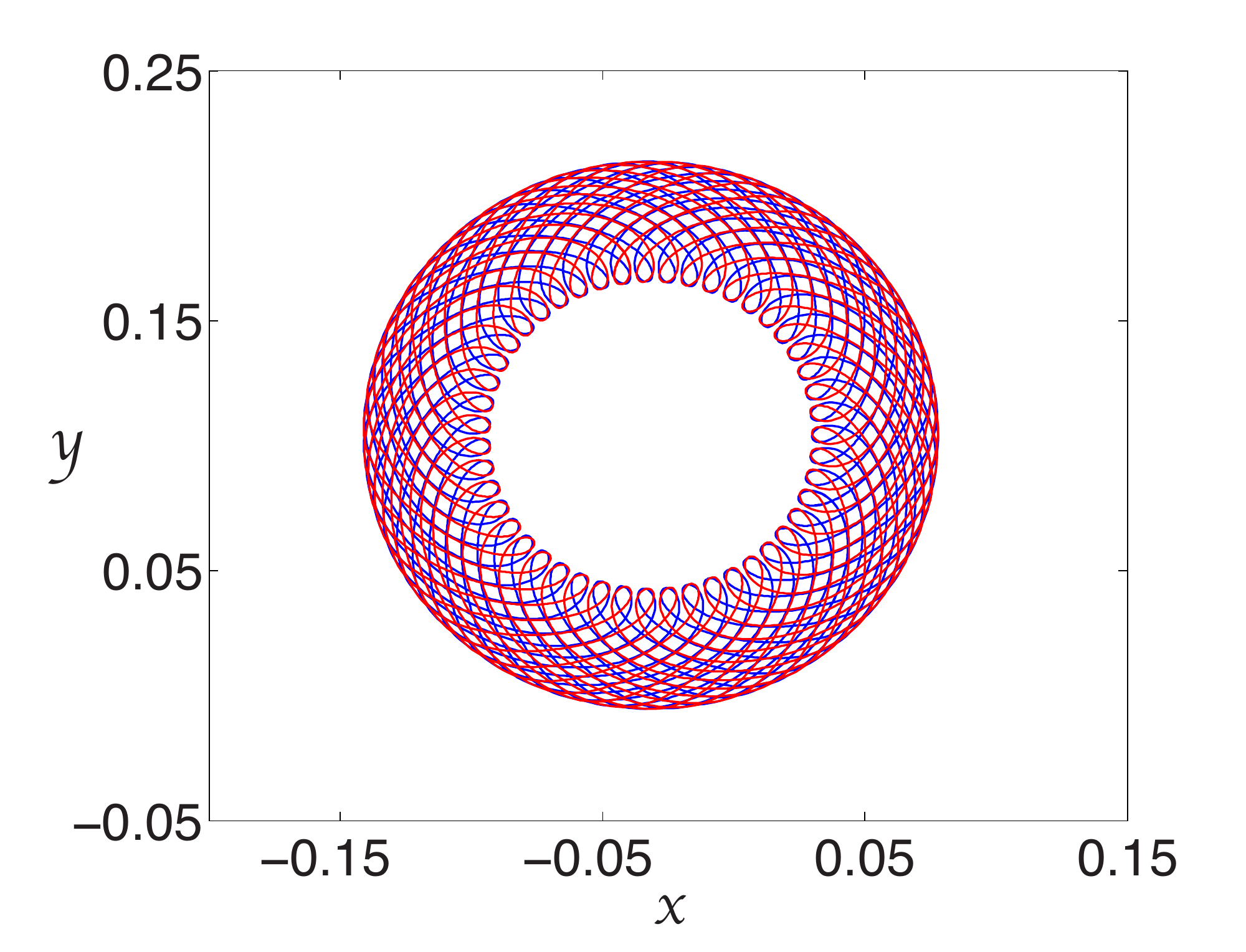}}
\end{center}
  	\caption[]{Dancing of two dipoles. In all cases, $z_1(0)=0$, $\alpha_1(0)=0$ and $\alpha_2(0)=5\pi/6$. Value of $z_2(0)$ in each case: $($a$)$ $0.01952+0.2\ii$, $($b$)$ $0.11+0.13007\ii$, $($c$)$ $0.08+0.13989\ii$, $($d$)$ $0.11+0.11601\ii$, $($e$)$ $0.04999+0.2009\ii$, $($f$)$ $0.01002+0.2\ii$. Arrows are only drawn in $($a$)$ and $($b$)$. The dipoles are tracing these trajectories in anti-clockwise direction.}
	\label{fig:2dipolesdancing}
\end{figure}

\paragraph{Synchronization.}
The most remarkable interaction mode of the two dipoles is the synchronization mode. We use the term synchronization to denote periodic trajectories where the two dipoles oscillate and interact with each other and return to the same position periodically. These periodic trajectories are found numerically using an iterative method. In particular, we use a shooting method that adjusts the initial position of one of the dipoles iteratively to hone in on the periodic orbit.

We distinguish two types of synchronization trajectories: {\em unbounded} and {\em bounded}. For the unbounded mode, the dipoles move side-by-side along undulating paths that exit the doubly-periodic domain to re-enter on the other side as depicted in Figure~\ref{fig:2dipolessynchronization}. In the bounded mode, the two dipoles dance around each other tracing out flower like orbits as depicted in Figure~\ref{fig:2dipolesdancing}.  These dancing trajectories are bounded in the sense that the dipoles move within  a confined region of the doubly-periodic domain.


\section{Stability of Rectangular and Diamond Lattices}
\label{sec:formation}

\begin{figure}[ft]
\begin{center}
 \subfigure[Rectangular lattice]{\includegraphics[scale=0.5]{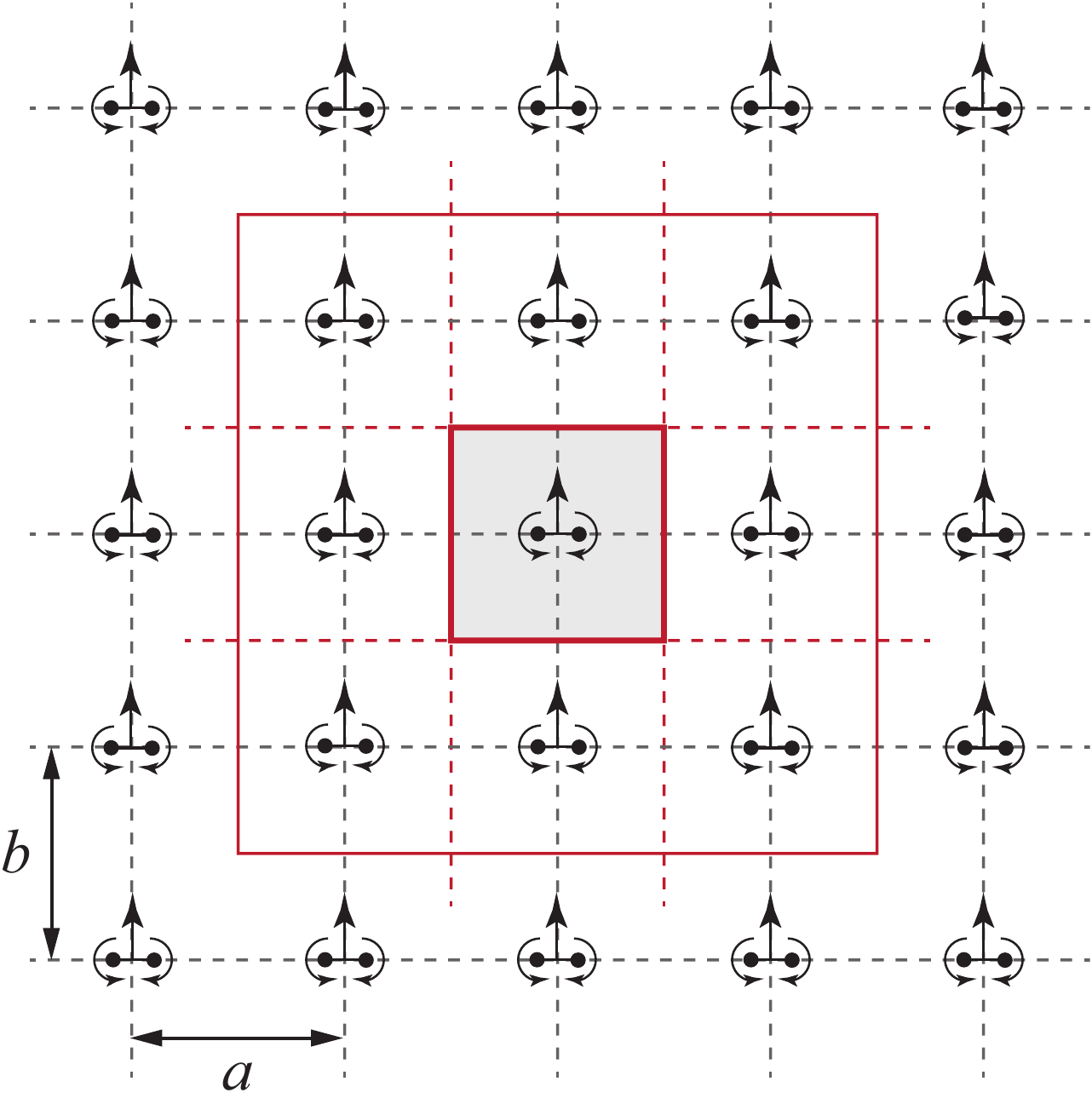}} \hspace{0.5in}
 \subfigure[Diamond lattice]{\includegraphics[scale=0.5]{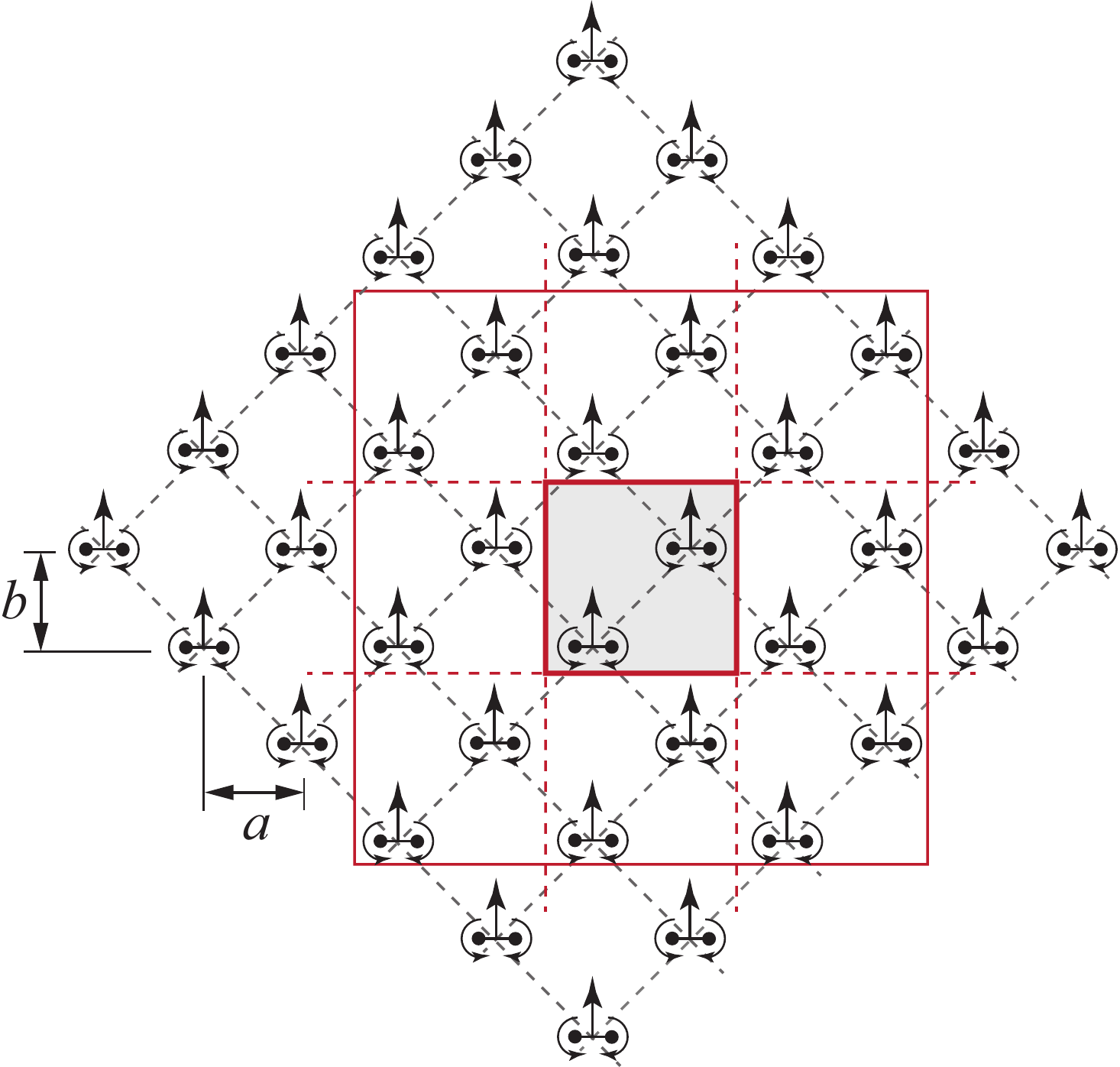}} 
\end{center}
  	\caption[]{Schematic of $($a$)$ rectangular lattice and $($b$)$ diamond lattice. The smallest doubly-periodic domain that generates the lattice is depicted in light grey.  A larger doubly-periodic domain or ``cell"  is also depicted in solid red. }
	\label{fig:Lattice}
\end{figure}

We address the dynamics and stability of two families of dipole lattices: rectangular and diamond (see  Figure~\ref{fig:Lattice}). By lattice, we mean an arrangement of dipoles in an ordered pattern extending to infinity in the unbounded plane. A {\em rectangular} lattice consists of dipoles aligned along $\alpha_{k,m} = \pi/2$ with their centers placed at $z_{k,m} = k a + \ii m b$, where $k,m= 0, \pm 1, \pm 2, \pm 3, \ldots$, $a$ denotes the distance
between two neighboring dipoles of the same row, and $b$ denotes
the distance between two rows. A {\em diamond} lattice consists of dipoles aligned along $\alpha_{k,m} = \pi/2$ but with centers placed such that $z_{k,m} = k a + i m b$, for $k,m$ even, and $z_{k,m} = (k+\dfrac{1}{2}) a + \ii (m+\dfrac{1}{2}) b$, for $k,m$ odd.
An alternative, and perhaps more elegant way, of describing these lattices is by considering them as special cases of dipoles in doubly-periodic domains. We adopt the latter view in this section. In particular, we define the `smallest' doubly-periodic domain (or `smallest cell') needed to describe these rectangular and diamond lattices. We then use the formulation in sections~\ref{sec:formulation} and~\ref{sec:singledipole} to prove that these configurations correspond to relative equilibria of the finite-dipole dynamical system and we analyze their linear stability. 

The `smallest' doubly-periodic domain needed to generate the rectangular lattice has half-periods $\omega_1=a/2$ and $\omega_2 = \ii b/2$ and contains a single dipole with orientation $\alpha(0)=\pi/2$,  see Figure~\ref{fig:Lattice}(a). The dipole's center is placed at the center of the domain for convenience. One can readily verify, using equation~\eqref{eq:formulation:eomOneDipole}, that, for all time, the dipole's orientation $\alpha$ remains unchanged while its center moves with constant velocity, thus the rectangular lattice is in a state of relative equilibrium. 

To generate the diamond lattice, the `smallest' doubly-periodic domain has half-periods $\omega_1=a/2$ and $\omega_2=\ii b/2$ and contains two finite dipoles $(z_1,\alpha_1)$ and $(z_2,\alpha_2)$ of equal strength $\Gamma$ and equal length $\ell_1 = \ell_2 = \ell$, with orientations $\alpha_1(0) = \alpha_2(0) = \pi/2$ and positions $z_2(0) = z_1(0) + (a/2 + \ii b/2)$, see Figure~\ref{fig:Lattice}(b).  To prove that this configuration is a relative equilibrium of the equations of motion, one needs to show that,  for all time, $\dot{\bar{z}}_1=\dot{\bar{z}}_2$ is constant and $\dot{\alpha}_1=\dot{\alpha}_2= 0$.  Actually, it suffices to show that $\dot{\conj{z}}_2-\dot{\conj{z}}_1=0$ and $\dot{\alpha}_1=\dot{\alpha}_2= 0$ for all time. Given these conditions, it immediately follows from \eqref{eq:formulation:eomTwoDipoles} that the lattice's velocity $\dot{\bar{z}}_1=\dot{\bar{z}}_2$ is constant for all time. 
The fact that the relative velocity $\dot{\conj{z}}_2-\dot{\conj{z}}_1=0$ is zero is obtained using~\eqref{eq:formulation:eomTwoDipoles} (and its analog for $\dot{\conj{z}}_2$) to get
\begin{equation}
\label{eq:diamondequil}
\begin{split}
\dot{\conj{z}}_2 -\dot{\conj{z}}_1& = \dfrac{\Gamma}{4\pi\ii}\Bigl[ \zeta(z_{2,\textrm{l}}-z_{1,\textrm{l}}) 
						- \zeta(z_{2,\textrm{l}}-z_{1,\textrm{r}}) 
						+ \zeta(z_{2,\textrm{r}}-z_{1,\textrm{l}}) 
						- \zeta(z_{2,\textrm{r}}-z_{1,\textrm{r}}) \\[2ex]
						& \hspace{0.75in} - \zeta(z_{1,\textrm{l}}-z_{2,\textrm{l}}) 
						+ \zeta(z_{1,\textrm{l}}-z_{2,\textrm{r}}) 
						- \zeta(z_{1,\textrm{r}}-z_{2,\textrm{l}}) 
						+ \zeta(z_{1,\textrm{r}}-z_{2,\textrm{r}}) \Bigr].
\end{split}
\end{equation}
Now, recall that the $\zeta$-function is an odd function (that is to say, $\zeta(z) = - \zeta(-z)$) and note that, for the diamond lattice, one has $z_2-z_1=z_{2,l}-z_{1,l}=z_{2,r}-z_{1,r}=\omega_1+\omega_2$. Then, it follows immediately from~\eqref{eq:diamondequil} that $\dot{\conj{z}}_2 -\dot{\conj{z}}_1=0$.
Similarly, to show that the dipoles' orientation is constant for all time, that is to say, that $\dot{\alpha}_1=\dot{\alpha}_2= 0$, 
rewrite~\eqref{eq:formulation:eomTwoDipolesAngle} in the form
\begin{equation}
\label{eq:diamondequil_angle}
\begin{split}						
	  \dot{\alpha}_1 & =\text{Re} \Bigl[ \frac{\Gamma}{4\pi\ii}\eee{\ii \alpha_{1}}\Bigl( \zeta(z_1-z_2-\frac{\ii\ell}{2}(\eee{\ii\alpha_1}+\eee{\ii\alpha_2})) +  
	  \zeta(z_1-z_2+\frac{\ii\ell}{2}(\eee{\ii\alpha_1}+\eee{\ii\alpha_2}))	
	  - 2\zeta(z_1-z_2) \Bigr)\Bigr].
\end{split}
\end{equation}
Now, recall the periodicity property of the $\zeta$-function and the fact that, for the diamond lattice $z_1-z_2 = - \omega_1 - \omega_2$, one can readily verify that the following identities hold
\begin{equation}
\label{eq:diamondperiodic}
\begin{split}
\zeta(z_1-z_2)&=\zeta\left(-\omega_1-\omega_2\right)=-\eta_1-\eta_2, \qquad \zeta(z_1-z_2+\ii\ell\eee{\ii\alpha})=\zeta(-\omega_1-\omega_2+\ii\ell\eee{\ii\alpha}), \\[2ex]
 &\zeta(z_1-z_2-\ii\ell\eee{\ii\alpha})=-\zeta(-\omega_1-\omega_2+\ii\ell\eee{\ii\alpha})-2 \eta_1-2 \eta_2. \\
\end{split}
\end{equation}
Substitute \eqref{eq:diamondperiodic} into~\eqref{eq:diamondequil_angle} and use the fact that  $\alpha_1(0)=\alpha_2(0)$ and $z_1-z_2$ is constant to get that $\dot{\alpha}_1(t)=\dot{\alpha}_1(0)=0$. The same result holds for $\dot{\alpha}_2$.
Thus, the diamond lattice is a relative equilibrium of the finite-dipole dynamical system. 

\begin{figure}[!t]
\begin{center}
 	\includegraphics[scale=0.45]{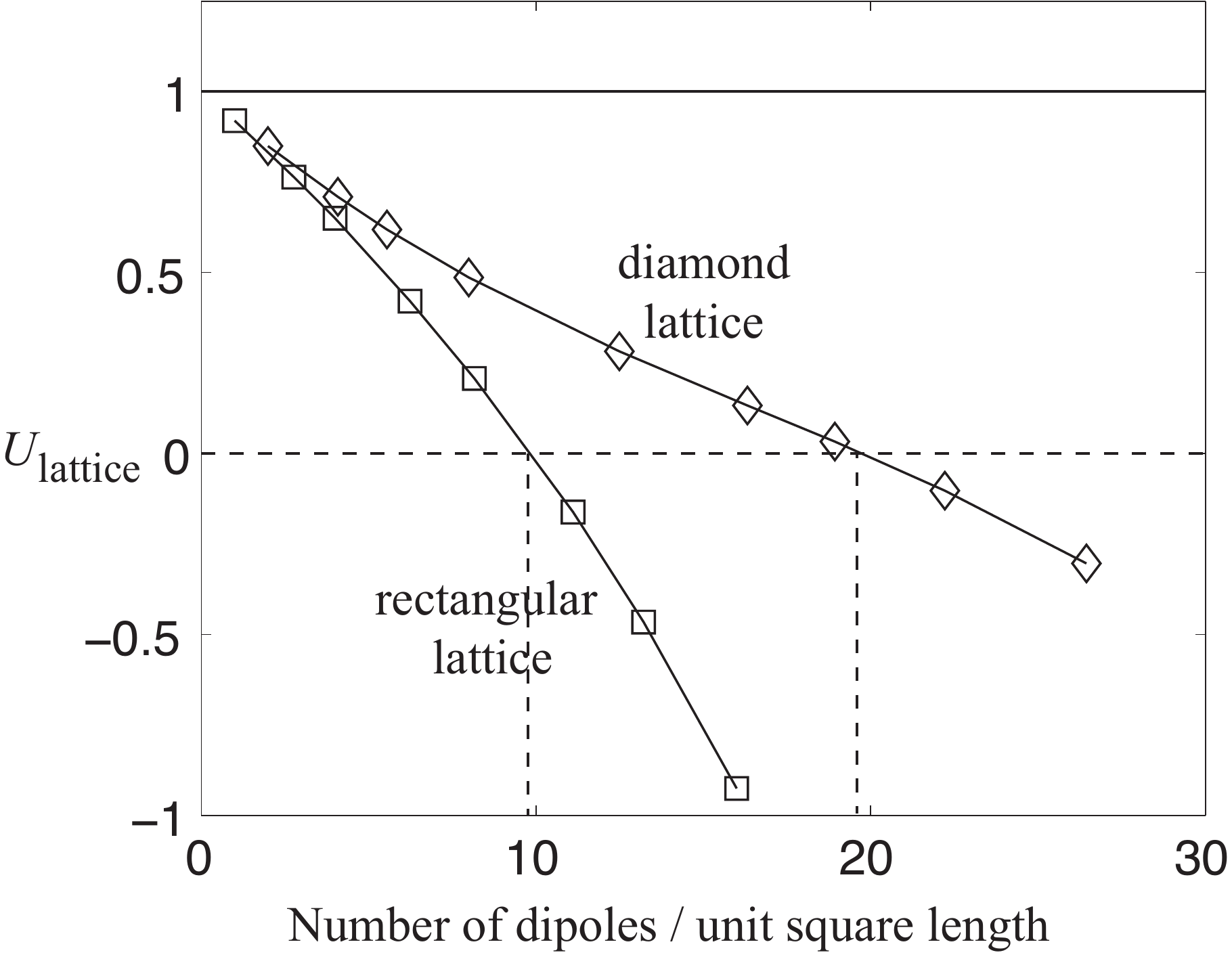}
\end{center}
  	\caption[]{Lattice translational velocity $U_{\rm lattice}$ versus its density for a square doubly-periodic domain $|\omega_1|=|\omega_2|=\omega$, and parameter values $\Gamma =1$, $\ell = 1/2\pi$.  Square symbol corresponds to the rectangular lattice while diamond symbol corresponds to the diamond lattice. A single dipole in an unbounded domain has unit velocity, shown in straight solid line.}
	\label{fig:dipolevel}
\end{figure}

Let $U_{\rm lattice}$ be the constant translational velocity of the lattice. It is instructive to compare the lattice velocity of the diamond and rectangular configurations with the velocity of a single dipole in an unbounded domain. For $\Gamma =1$ and $\ell = 1/2\pi$, one can readily see using~\eqref{eq:formulation:wnsi} that the velocity of a single dipole in an unbounded domain is equal to $1$. We use the same parameter values and compute $U_{\rm lattice}$ from~\eqref{eq:formulation:eomOneDipole} and~\eqref{eq:formulation:eomTwoDipoles} for the rectangular and diamond lattices, respectively. The results are shown in Figure~\ref{fig:dipolevel} as a function of the lattice density, defined as the number of dipoles per unit square length. Obviously, neither the rectangular nor the diamond formation present any advantages over the single dipole in term of increased translational velocity. To the contrary, the hydrodynamic interactions cause the dipoles in these lattices to move slower than the single dipole with the rectangular lattice being slowest for all shown densities. A further increase in the lattice density causes $U_{\rm lattice}$ to reverse sign and the dipoles to move in the direction opposite to their self-induced velocity. By continuity arguments, one deduces that there exist critical density values for which the rectangular and diamond lattices are stationary.

We now examine the linear stability of these relative equilibria. Typically, the stability of infinite lattices is analyzed by introducing infinitesimal perturbations on each dipole's position and orientation, and  looking for plane wave solutions of the linearized system; see, for example,~\cite{saffman:1993a} for a review of the  stability analysis of a row of point vortices and of a von K\'{a}rm\'{a}n street. See also \cite{tkachenko:sjetp1966b} for stability of 2D vortex lattices and the more recent work \cite{desreumaux:epje2012a} on the stability of driven and motile particle lattices in confined geometry.  This approach involves infinite sums whose convergence needs to be established. Here, we avoided this complication by using a doubly-periodic domain and the Weierstrass $\zeta$-function. Indeed, in~\cite{aref:jfm1995a}, Aref showed that the stability of an infinite row of point vortices can be formulated and studied as the stability of point vortices in a periodic domain. He noted that the perturbation wave solution is equivalent to the eigenvalue problem associated with point vortices in a periodic domain. Further, since a wave of any wavelength must repeat after a finite number of vortices, various wavelengths can be captured by considering vortices in a periodic domain of various ``cell" sizes. A cell is a doubly-periodic domain that is not necessarily the smallest, as depicted in Figure~\ref{fig:Lattice}. We follow Aref's approach in the sense that we consider dipoles in a doubly-periodic domain and we apply perturbations in cells of various sizes to analyze how the stability of the lattice depends on the periodicity of the perturbation.

 For concreteness, we consider a doubly-periodic cell containing $N$ dipoles. Let $\delta z_n$ and  $\delta \alpha_n$, $n=1,2,\ldots,N$, denote the infinitesimal perturbations on the position and orientation of each dipole in this cell so that
 $z_n = U_{\rm lattice} + \delta z_n$ and $\alpha_n = \dfrac{\pi}{2} +\delta \alpha_n$. 
 Due to the doubly-periodic nature of the problem, these perturbations will be repeated periodically. We linearize equations \eqref{eq:formulation:eomZn} and \eqref{eq:formulation:eomAlphaN} about the unperturbed lattice configuration and make use  
of the formula $d\zeta(z)/dz=-\rho(z)$, with $\rho(z)$ being the Weierstrass Elliptic function defined as    
\begin{equation}
\label{eq:WeierstrassEllipticFunction}
 	\rho\left(z;\omega_{1},\omega_{2}\right)=\frac{1}{z^2}+\sum_{p,q} \frac{1}{(z-\Omega_{pq})^2} -\frac{1}{\Omega_{pq}^{2}},
	\qquad
	p,q\in \mathbb{Z} \!-\!\{0\}. 
\end{equation}
The linearized perturbed equations can be written in matrix form as follows
\begin{equation}
\dfrac{d}{dt}\left(\begin{array}{c} \delta x_i \\ \delta y_i \\ \delta\alpha_i \end{array}\right) = M_{ij} \left(\begin{array}{c} \delta x_j \\ \delta y_j \\ \delta \alpha_j \end{array} \right), \qquad i,j = 1,\ldots,N
\end{equation}
The eigenvalues of the $M_{ij}$ matrix are computed numerically for all $i,j$. The lattice is said to be linearly stable to a given perturbation if all eigenvalues of $M_{ij}$ have non-positive real parts, that is to say, if all Re($\lambda$)$\leq 0$. The analysis is performed systematically by considering doubly-periodic cells of various sizes, starting with the smallest domain size $a$, $b$. When the cell size is equal to the smallest domain, the same perturbation is applied to all dipoles. The values of the largest Re($\lambda$) are tabulated in Table 1 for the rectangular lattice and Tables 2 and 3 for the diamond lattice.

\begin{table}
\caption{Rectangular lattice.}
\begin{center}
\begin{tabular}{ll|rrrlrr}
\hline
& & \multicolumn{5}{c}{largest Re$(\lambda)$}    \\
\hline 
& & \multicolumn{3}{c}{$a=1$}\vline &\multicolumn{2}{c}{$b=1$}  \\
\hline
Cell Size    & N & \multicolumn{1}{c}{$b=1$}& \multicolumn{1}{c} {$b=1.5$}& \multicolumn{1}{c}{$b=2$} \vline & \multicolumn{1}{c}{$a=1.5$} & \multicolumn{1}{c} {$a=2$}\\
\hline
$a,b$     & $1$    & \multicolumn{1}{c} {$0$}   &\multicolumn{1}{c} {$0$}	&  \multicolumn{1}{c} {$0$} \vline &  \multicolumn{1}{c} {$0$} & \multicolumn{1}{c}{$0$}		\\ 
$2a,2b$ & $4$   & \multicolumn{1}{c} {$1.09$}  & \multicolumn{1}{c} {$0.84$} & \multicolumn{1}{c} {$0.80$} \vline	&\multicolumn{1}{c}{$0.84$}  &\multicolumn{1}{c} {$0.80$} \\
$3a,3b$ & $9$  & \multicolumn{1}{c} {$0.91$}   &\multicolumn{1}{c} {$0.73$}	& \multicolumn{1}{c} {$0.71$}	\vline &\multicolumn{1}{c}{$0.73$}  &\multicolumn{1}{c} {$0.70$}	\\
$4a,4b$ & $16$  & \multicolumn{1}{c} {$1.09$}   &\multicolumn{1}{c} {$0.84$} & \multicolumn{1}{c} {$0.80$}	\vline &\multicolumn{1}{c}{$0.84$}  &\multicolumn{1}{c} {$0.80$}	\\
$5a,5b$ & $25$  & \multicolumn{1}{c} {$1.02$}  &\multicolumn{1}{c} {$0.80$}	& \multicolumn{1}{c} {$0.77$}\vline	&\multicolumn{1}{c} {$0.80$}  &\multicolumn{1}{c} {$0.76$}\\
$6a,6b$ & $36$  & \multicolumn{1}{c} {$1.09$}  &\multicolumn{1}{c} {$0.84$}	& \multicolumn{1}{c} {$0.80$}	\vline &\multicolumn{1}{c}{$0.84$} 	& \multicolumn{1}{c} {$0.80$}\\
\hline
\end{tabular}
\end{center}
\end{table}

This analysis shows that the rectangular lattice is always unstable while the diamond lattice can be either unstable or linearly stable, depending on the lattice parameters $a$ and $b$ and on the size of the cell where the perturbation is applied. For example, when the 
same perturbation is applied to all dipoles, that is to say, when the size of the cell where the perturbation is applied is the same as the size of the smallest doubly-periodic domain, both the rectangular and diamond lattices are linearly stable. Also, when $a=b=1$, the diamond lattice is always linearly stable but not the rectangular lattice. For $a=2$, $b=1$, the diamond  lattice is always unstable (except as we just noted when the perturbation domain is the smallest doubly-periodic domain).

\begin{figure}[!t]
\begin{center}
 	\subfigure[at t = 0]{\includegraphics[scale=0.4]{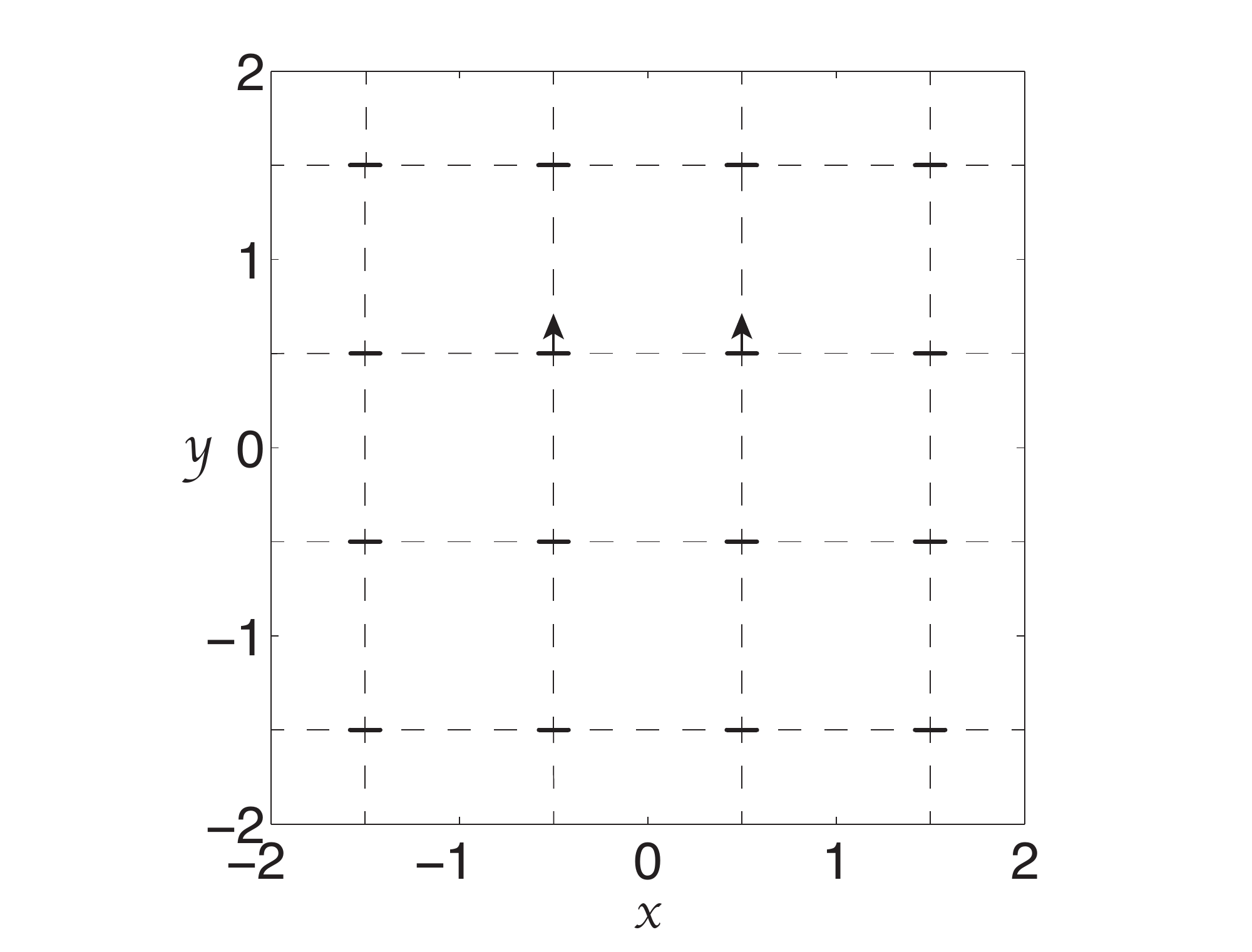}} 
	\subfigure[at t = 3]{\includegraphics[scale=0.4]{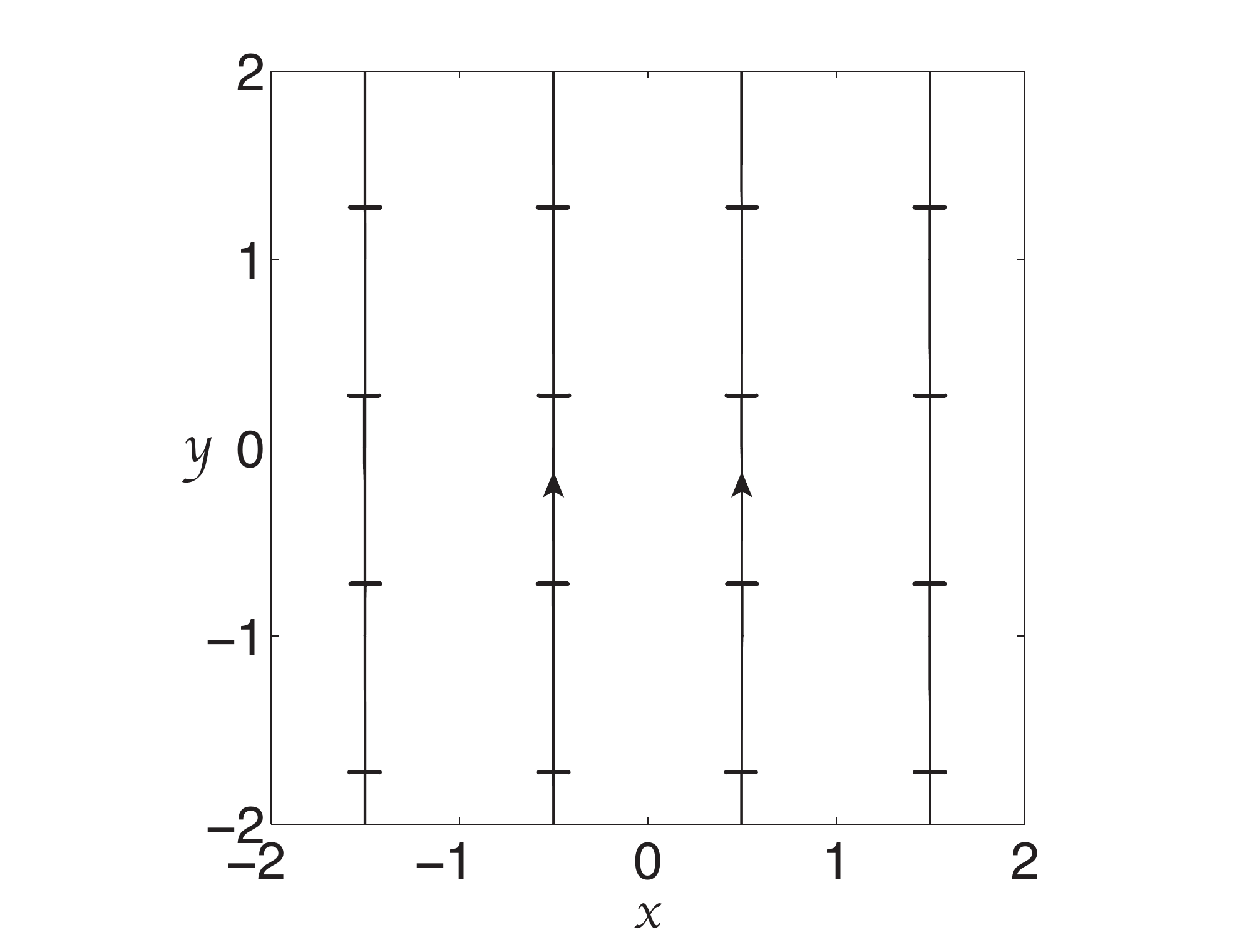}}
	\subfigure[at t = 6.5]{\includegraphics[scale=0.4]{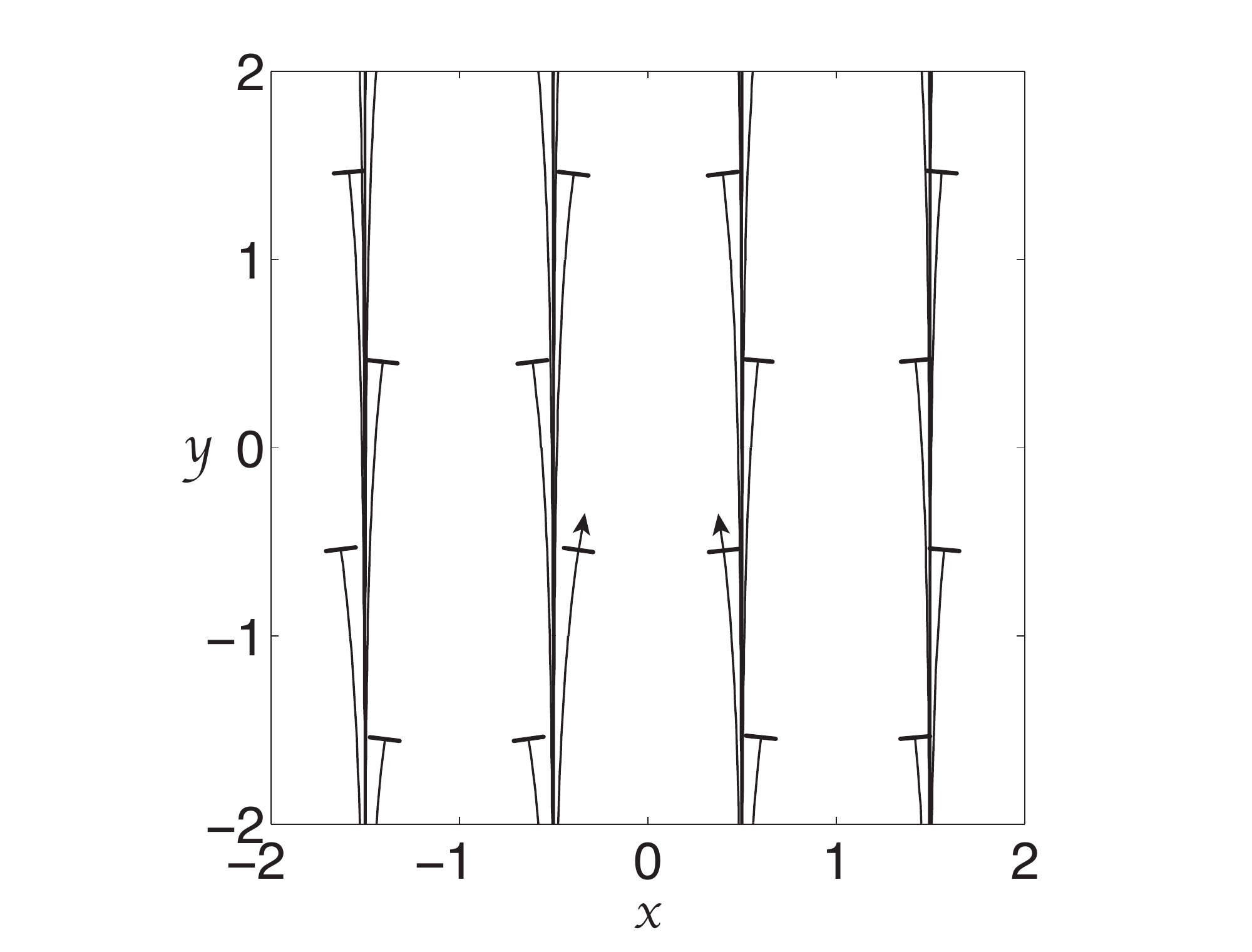}}
	\subfigure[at t = 10]{\includegraphics[scale=0.4]{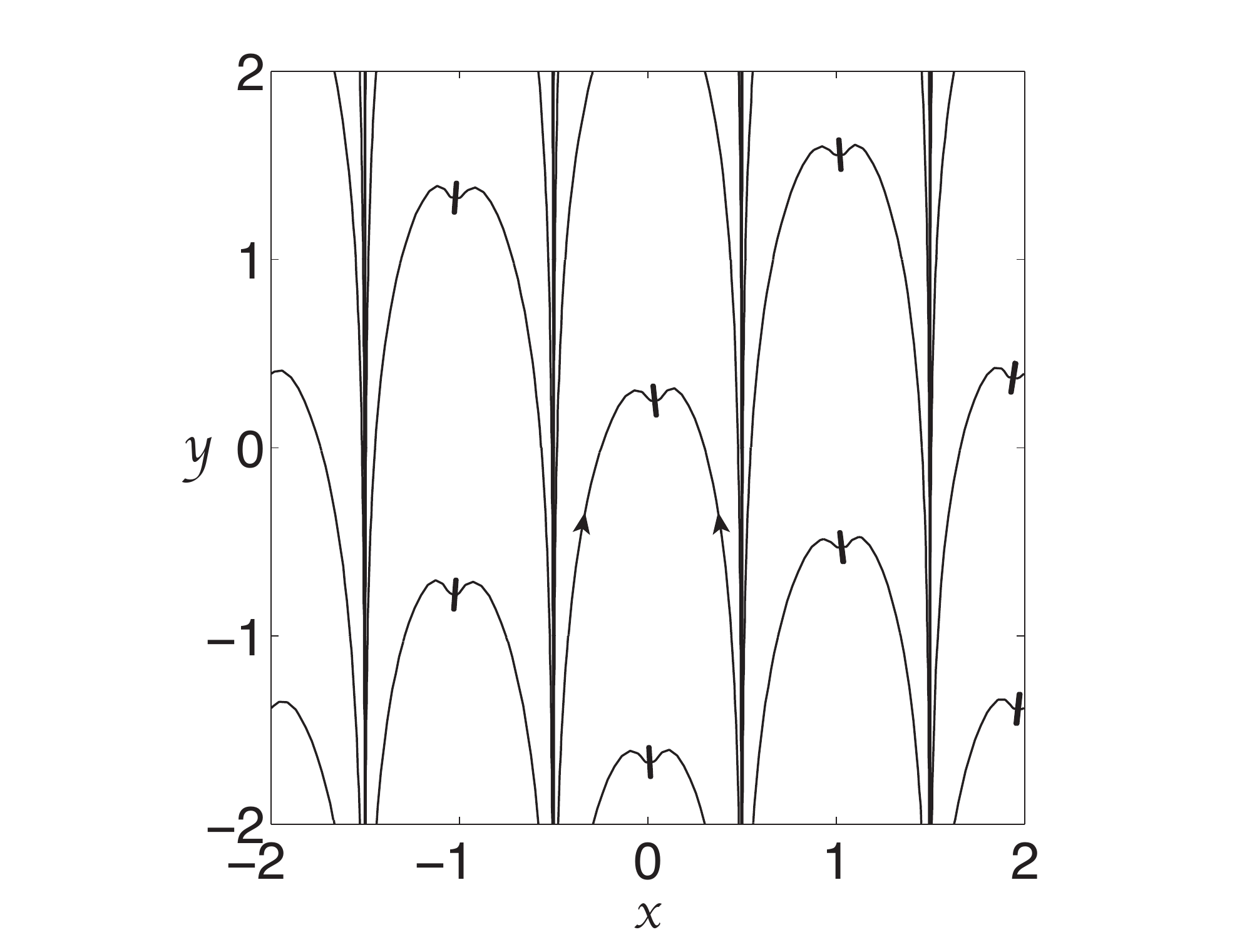}}
\end{center}
  	\caption[]{Time evolution of rectangular lattice for parameter values $a= b =1$, $\Gamma =1$ and $\ell = 1/2\pi$. $($a$)$ initial configuration 
	of rectangular lattice subject to random perturbations in the shown cell ($N=16$). $($b$)$ and $($c$)$ trajectories of the dipoles at two different times. $($d$)$ Collision of dipoles and break down of the dipole lattice.}
	\label{fig:rectFormation}
\end{figure}

\begin{figure}[!t]
\begin{center}
 	\subfigure[at t = 0]{\includegraphics[scale=0.4]{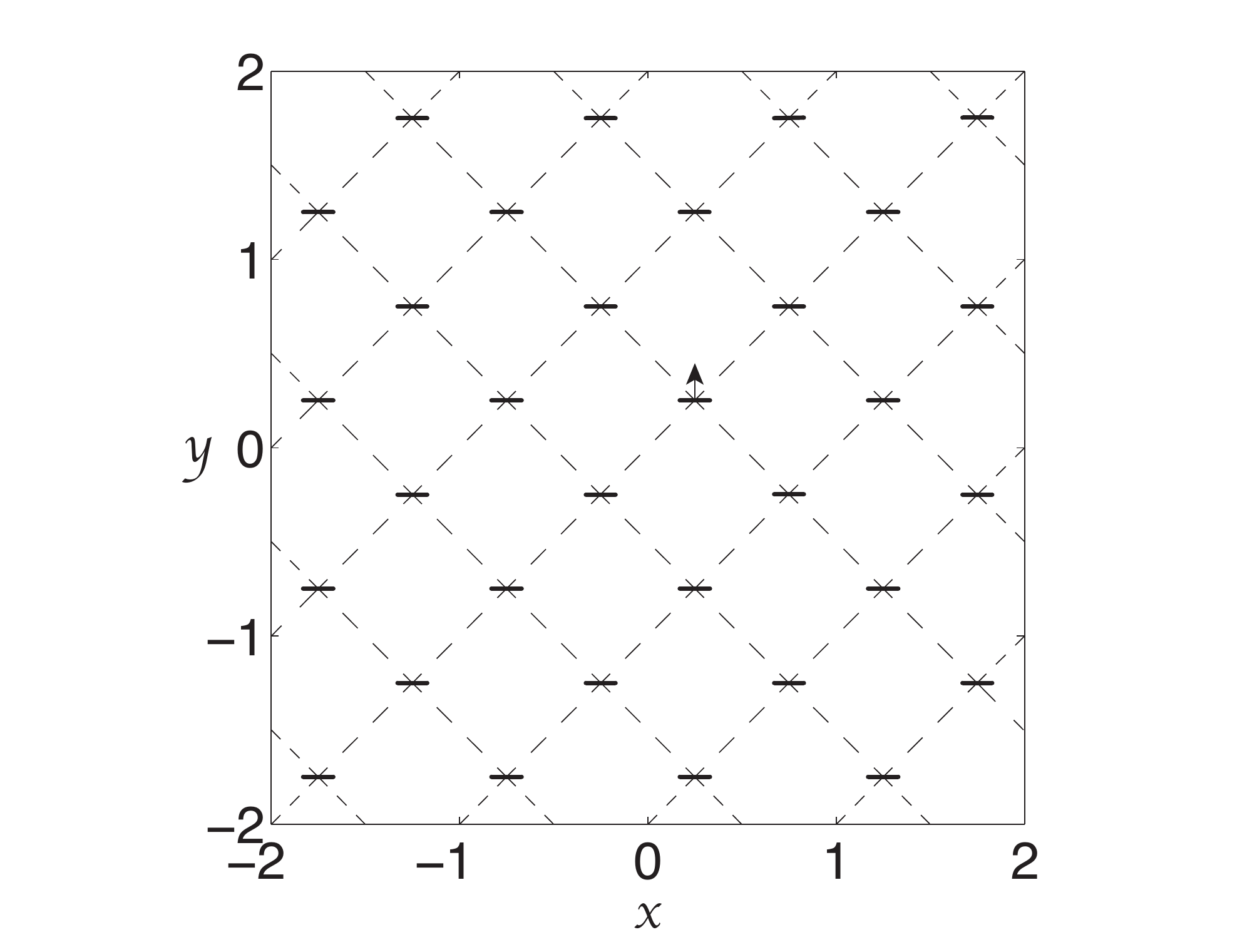}} 
	\subfigure[at t = 30]{\includegraphics[scale=0.4]{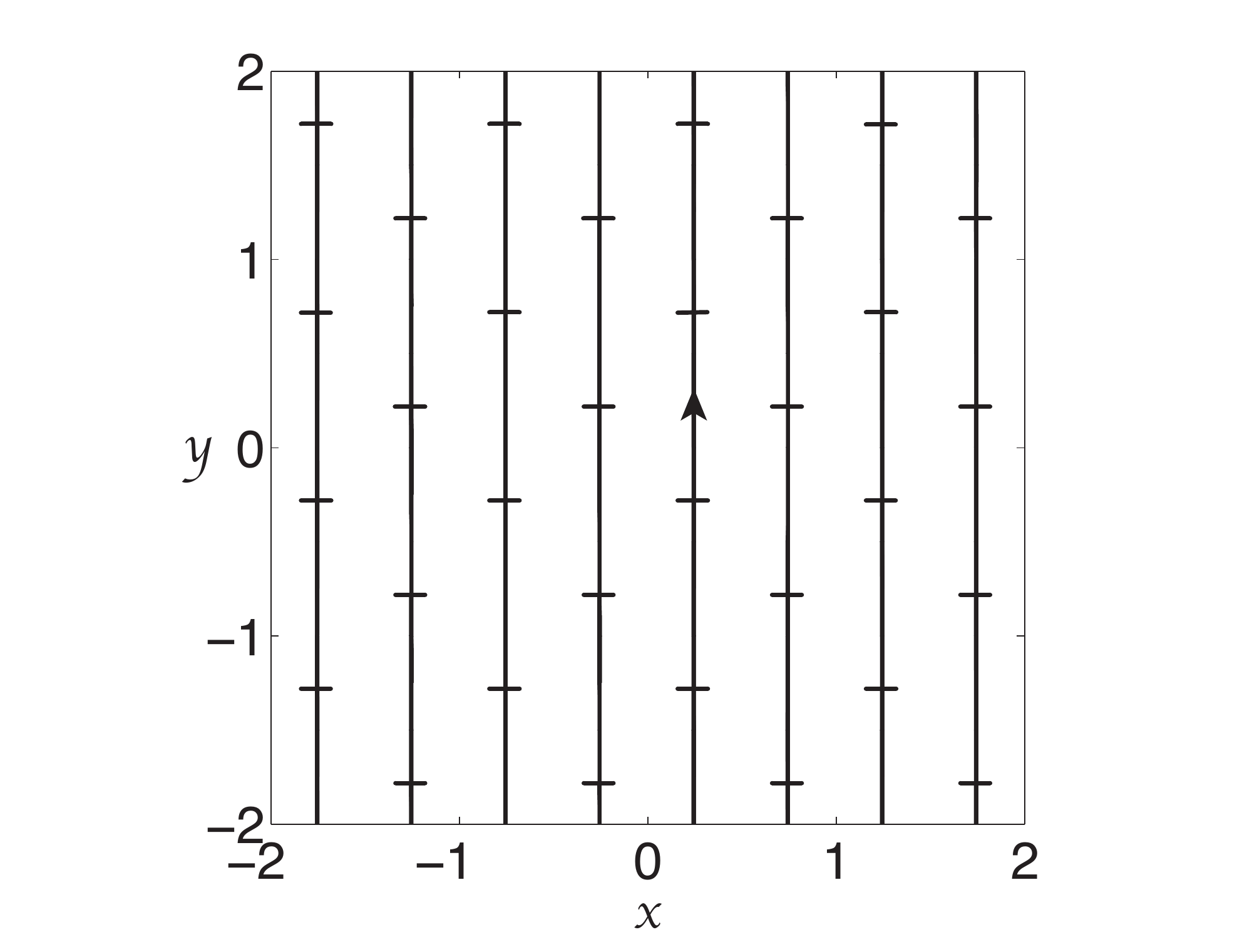}}
\end{center}
  	\caption[]{Time evolution of diamond lattice for parameter values $a= b =1$, $\Gamma =1$ and $\ell = 1/2\pi$. (a) initial configuration 
	of diamond lattice subject to random perturbations in the shown cell ($N=32$). (b) trajectories of the dipoles after an integration time $T =30$ time units.}
	\label{fig:diamondFormation}
\end{figure}

These results have been confirmed by numerically integrating the nonlinear equations in~\eqref{eq:formulation:eomZn} and~\eqref{eq:formulation:eomAlphaN} for the perturbed lattices. The perturbations are chosen randomly such that their magnitude is of the order $a/1000$. For the cases predicted to be unstable by the eigenvalue analysis, the lattices break down in finite time. Figure~\ref{fig:rectFormation} provides snapshots of the collapse of a rectangular lattice subject to initial random perturbations applied in the shown domain ($N =16$). Meanwhile, for the linearly stable cases, the lattice keeps its integrity as shown in Figure~\ref{fig:diamondFormation} for a diamond lattice with parameters $a=b=1$. The diamond formation persisted to the end of the integration time ($T=100$ time units).

\begin{table}
\caption{Diamond lattice $(b=1)$}
\begin{center}
\begin{tabular}{ll|rrrrrrr}
\hline
& & \multicolumn{7}{c}{largest Re$(\lambda)$}    \\
\hline
Cell Size    & N & $a = 1$  & $a = 1.1$ & $a=1.2$ & $a=1.3$ & $a=1.4$ & $a=1.5$ & $a=2$ \\
\hline
$a,b$     & $2$    & $0$  & $0$     &	$0$ &	$0$ &	 $0$ &	$0$ &	$0$\\ 
$2a,2b$ & $8$   & $0$   & $0$   &	$0$ &	$0$ &	$0$ &	$0$ &	$0.67$\\
$3a,3b$ & $18$  & $0$   & $0$   &	$0$ &	$0$ &	$0$ &	$0.15$ 	&	$0.58$	\\
$4a,4b$ & $32$  & $0$   &$0$    &	$0$ &	$0$ &	$0$ &	$0.15$	&	$0.67$\\
$5a,5b$ & $50$  & $0$  & $0$   &	$0$ &	$0$ &	$0.07$	&	$0.13$	&	$0.64$\\
$6a,6b$ & $72$  & $0$  & $0$   &   $0$ &	$0$ &	$0.09$	&	$0.15$ 	&	$0.67$\\
\hline
\end{tabular}
\end{center}
\end{table}

\begin{table}
\caption{Diamond lattice $(a=1)$}
\begin{center}
\begin{tabular}{ll|rrrrr}
\hline
& & \multicolumn{5}{c}{largest Re$(\lambda)$}    \\
\hline
Cell Size    & N  & $b = 1.1$ & $b=1.2$ & $b=1.3$ & $b=1.4$ & $b=1.5$ \\
\hline
$a,b$     & $2$     &	$0$	& 	$0$	&	$0$	&	$0$	&	$0$\\ 
$2a,2b$ & $8$    &    	$0$	&  	$0$	&	$0$	&	$0.16$	&	$0.42$\\
$3a,3b$ & $18$  &  	$0$	&    	$0$	&	$0$	&	$0.22$	&	$0.37$\\
$4a,4b$ & $32$  &   	$0$	&   	$0$	&	$0$	&	$0.30$	&	$0.42$\\
$5a,5b$ & $50$  &   	$0$	&	$0$	&	$0.18$	&	$0.33$	&	$0.42$\\
$6a,6b$ & $72$  & 	$0$	& 	$0$	&	$0.20$	&	$0.34$	&	$0.45$\\
\hline
\end{tabular}
\end{center}
\end{table}

\begin{figure}[!t]
\begin{center}
 	\includegraphics[scale=0.33]{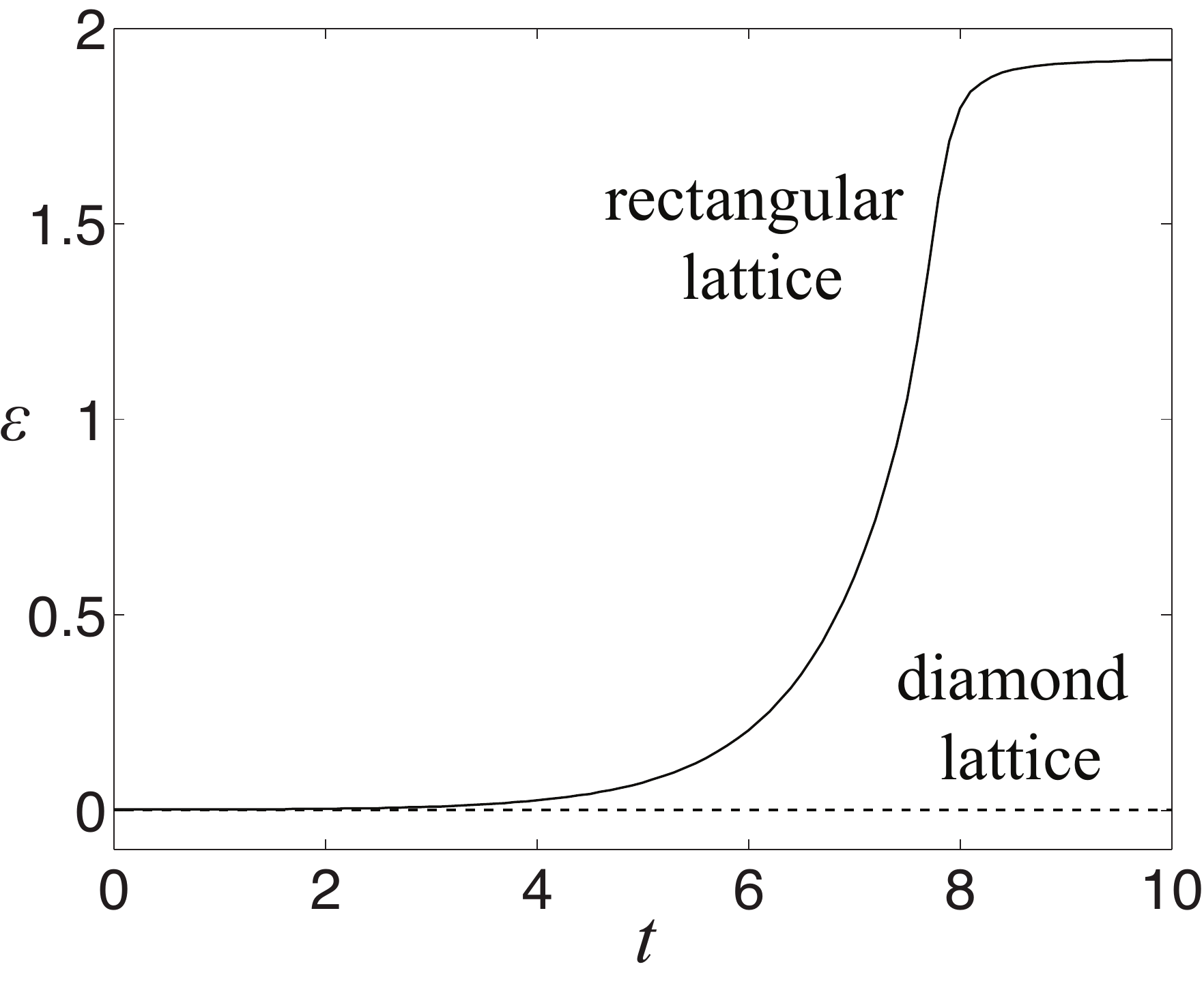}
\end{center}
  	\caption[]{$\epsilon$ versus time $t$ where $\epsilon$  is the deviation of the perturbed lattices in Figures~\ref{fig:rectFormation} and \ref{fig:diamondFormation} from their respective unperturbed structure. Clearly, the rectangular lattice looses its lattice structure whereas the diamond lattice maintains its lattice integrity.
	}
	\label{fig:formationerror}
\end{figure}

To quantify the deviation from the unperturbed lattice structure, we compare the dipoles positions at each time $t$ with that of the unperturbed lattices using
\begin{equation}
	\label{eq:errorparameter}
	\epsilon(t)=\frac{\sum_{p,q}\abs{\abs{z_{p}(t)-z_{q}(t)}^2-\abs{z_{p}^{\rm lattice}-z_{q}^{\rm lattice}}^2}}{(N-1)(N-2)/2}, \qquad p\neq q.
\end{equation}
This expression can be thought of as the mean square deviation of the perturbed lattice compared to the unperturbed lattice and its value is shown in Figure~\ref{fig:formationerror} for the perturbed rectangular and diamond lattices of Figures~\ref{fig:rectFormation} and \ref{fig:diamondFormation}, respectively. Clearly, the deviation of the perturbed rectangular lattice begin to grow rapidly around $t = 3$ time units, while the deviation of the perturbed diamond lattice remains small for all integration time.

We conclude this section by commenting on the insights these results provide in the context of fish schooling.
For fish schools, it has been argued that the diamond formation is favorable from an energy efficiency standpoint,~\cite{weihs:n1973a}. In this seminal work, Weihs based his analysis on a stationary infinite diamond lattice and computed the locomotory benefits a given fish gets from the vortical wakes of neighboring fish. The wakes were modeled as idealized vortex streets and the fish were assumed to be point particles. That is to say, Weihs' model accounted for the near-field effect of fish wakes. The finite-dipole model considers the far-field hydrodynamic coupling (neglecting near-field vorticity) of self-propelled swimmers, and, as such, can be viewed as complementary to Weihs' model with the important difference that it allows for dynamic interactions among the fish (dipoles) whereas the latter assumes stationary fish. 
Based on the finite-dipole model, we make the following observations:
\begin{itemize}
\item[(i)] Neither the rectangular nor the diamond dipole lattices provide locomotory advantages to the individual dipoles in the sense that the lattice translational velocity is smaller than the velocity of a self-propelled dipole in an unbounded plane (see Figure~\ref{fig:dipolevel}). Perhaps not surprisingly, this result emphasizes that any locomotory advantages to schooling in terms of efficiency of motion would arise from near-field vortical wakes, as in Weihs' model, and not from far-field effects. Extraction of energy from near-field vorticity has been confirmed experimentally in live and dead trout, see~\cite{Liao:science2003a,Beal:jfm2005a}. 

\item[(ii)]  However, our model shows that the diamond formation is beneficial from a stability standpoint. It is not clear how much stability is a desirable feature in large fish schools on the move. Stability, which measures how much a system opposes change, limits maneuverability. Here, we are referring to the stability and maneuverability of the school as opposed to that of  the individual fish, the latter has been the topic of several studies, see, for example,~\cite{weihs:icb2002a} and references therein. We conjecture that passive stability of the school when subject to small perturbations might be desirable to a migrating school of fish.  Active stabilization to stay in a school is energetically costly and therefore it may be more beneficial to travel in a school formation that is passively stable and requires no or little additional effort to maintain. These statements are yet to be validated by experimental observations. If true, they imply that the diamond formation is beneficial for both energy extraction from near-field wakes,~\cite{weihs:n1973a}, as well as for passive stabilization of the school formation when subject to small perturbations.
\end{itemize}


\section{Conclusions}

We derived equations of motion for a system of finite dipoles in a doubly-periodic domain. We started from the standard point vortex equations in doubly-periodic domains and followed an approach similar to that in~\cite{Tchieu:prsa2012a} for finite dipoles in unbounded plane. We used the resulting equations of motion to examine the motion of one and two dipoles in doubly-periodic domains. We showed that a single dipole in a doubly-periodic domain can exhibit periodic and aperiodic motion, whereas two dipoles exhibit a range of interesting behavior including collision, collision-avoidance, and motion synchronization. In the latter category, the two dipoles travel in synchrony along unbounded and bounded periodic trajectories due to hydrodynamic coupling only. In the context of fish schooling, our main motivation for considering this class of models, these trajectories imply that hydrodynamic interactions may be responsible, at least in part, for the remarkable synchrony of motion observed in schools of fish. Further, the bounded periodic trajectories reported here are reminiscent to the stable epicyclic orbits that were observed in the context of vortex dipoles in the dilute-gas regime of a Bose--Einstein condensate,~\cite{Middelkamp:pr2011a}. Indeed, it is known that equations governing quantized vortices in helium II and Bose--Einstein condensates are the same as that in ideal, incompressible fluids (see \cite{donnelly:1991a}). A formal connection between quantized vortices and the finite-dipole model is beyond the scope of the present paper.

We then identified two families of relative equilibria consisting of rectangular and diamond lattices, respectively. We examined the linear stability of these dipole lattices and found that the rectangular dipole lattice is always unstable whereas the diamond lattice is linearly stable for a range of parameter values and perturbation domains. In the context of fish schools, we argued that active stabilization to stay in the school formation is energetically costly and therefore it may be more beneficial for a migrating school of fish to travel in a diamond formation that is passively stable to small perturbations and requires no or little additional effort to maintain.

Finally, we note that, motivated by recent advances in microfluidics, the motion of self-propelled particles (bacteria) in two-dimensional fluid channels (Hele-Shaw cells) has been the topic of several recent studies; see, for example,~\cite{desreumaux:epje2012a} and references therein. It is well-known that the equations of motion governing Hele-Shaw flows, though viscosity-driven, are identical to those of the inviscid potential flow. Therefore, we expect our model to be applicable in the microfluidic context as well. This direction will be pursued in future work.

%
%

\paragraph{Acknowledgement.} This work is partially supported by the National Science Foundation through the CAREER award CMMI 06-44925 and grant CCF 08-11480.


\bibliographystyle{unsrt}
\bibliography{reference1}

\end{document}